\documentclass[sigconf]{acmart}

\usepackage{amsmath,amsfonts}

\usepackage{amsthm}
\usepackage{algorithm}
\usepackage{algpseudocode}

\usepackage{graphicx}
\usepackage{textcomp}
\usepackage{xcolor}
\usepackage{mdframed}
\usepackage{tabularx}
\usepackage{comment}
\usepackage{booktabs}
\usepackage{epsfig}
\usepackage{bm}
\usepackage[tight,footnotesize]{subfigure}
\usepackage{adjustbox}
\usepackage{multirow}
\usepackage{cleveref}

\newtheorem{theorem}{Theorem}

\renewcommand\footnotetextcopyrightpermission[1]{} 
\acmConference[Conference acronym 'XX]{Make sure to enter the correct
  conference title from your rights confirmation emai}{June 03--05,
  2018}{Woodstock, NY}

\begin{document}

\title{IPComp: \texorpdfstring{\underline{I}}{I}nterpolation Based \texorpdfstring{\underline{P}}{P}rogressive Lossy \texorpdfstring{\underline{Comp}}{Comp}ression for Scientific Applications}


\author{Zhuoxun Yang}
\affiliation{%
  \institution{Florida State University}
  \city{Tallahassee}
  \state{FL}
  \country{USA}}
\email{zy24b@fsu.edu}

\author{Sheng Di}
\affiliation{%
  \institution{The University of Chicago}
  \institution{Argonne National Laboratory}
  \city{Lemont}
  \state{IL}
  \country{USA}}
\email{sdi1@anl.gov}

\author{Longtao Zhang}
\affiliation{%
  \institution{Florida State University}
  \city{Tallahassee}
  \state{FL}
  \country{USA}}
\email{lzhang11@fsu.edu}

\author{Ruoyu Li}
\affiliation{%
  \institution{Florida State University}
  \city{Tallahassee}
  \state{FL}
  \country{USA}}
\email{rl13m@fsu.edu}

\author{Ximiao Li}
\affiliation{%
  \institution{Florida State University}
  \city{Tallahassee}
  \state{FL}
  \country{USA}}
\email{xl24g@fsu.edu}

\author{Jiajun Huang}
\affiliation{%
  \institution{University of California, Riverside}
  \city{Riverside}
  \state{CA}
  \country{USA}}
\email{jhuan380@ucr.edu}

\author{Jinyang Liu}
\affiliation{%
  \institution{University of Houston}
  \city{Houston}
  \state{TX}
  \country{USA}}
\email{jliu217@central.uh.edu}

\author{Franck Cappello}
\affiliation{%
  \institution{The University of Chicago}
  \institution{Argonne National Laboratory}
  \city{Lemont}
  \state{IL}
  \country{USA}}
\email{cappello@mcs.anl.gov}

\author{Kai Zhao}
\authornote{Corresponding author}
\affiliation{%
  \institution{Florida State University}
  \city{Tallahassee}
  \state{FL}
  \country{USA}}
\email{kzhao@cs.fsu.edu}



\begin{abstract}
Compression is a crucial solution for data reduction in modern scientific applications due to the exponential growth of data from simulations, experiments, and observations.
Compression with progressive retrieval capability allows users to quickly access coarse approximations of data and then incrementally refine these approximations to higher fidelity. 
Existing progressive compression solutions suffer from low reduction ratios or high operation costs, effectively undermining the approach's benefits.
In this paper, we propose the first-ever interpolation-based progressive lossy compression solution that has both high reduction ratios and low operation costs. The interpolation-based algorithm has been verified as one of the best for scientific data reduction, but previously, no effort exists to make it support progressive retrieval. Our contributions are three-fold:
(1) We thoroughly analyze the error characteristics of the interpolation algorithm and propose our solution, IPComp, with multi-level bitplane and predictive coding. 
(2) We derive optimized strategies toward minimum data retrieval under different fidelity levels indicated by users through error bounds and bitrates.
(3) We evaluate the proposed solution using six real-world datasets from four diverse domains. Experimental results demonstrate our solution archives up to $487\%$ higher compression ratios and $698\%$ faster speed than other state-of-the-art progressive compressors, and reduces the data volume for retrieval by up to $83\%$ compared to baselines under the same error bound, and reduces the error by up to $99\%$ under the same bitrate.

\end{abstract}

\maketitle  

\section{Introduction}
\label{sec:intro}

The increasing volume of scientific data generated by simulations, instruments, and observations has outpaced the data processing, storage, and transfer capabilities of modern computer systems, including both workstations and supercomputers. 
For example, in climate research, the Coupled Model Intercomparison Project (CMIP)~\cite{cmip} aims to advance our understanding of the climate system by coordinating standardized experiments with Earth System Models (ESMs) and thus enabling comprehensive comparisons of how different models represent past, present, and future climate conditions. Thanks to the rapid evolution of leading HPC systems, each successive phase of CMIP has seen significant increases in data volume -- CMIP3 generated around 40 TB~\cite{cmip-surrogate, cmip6-plan}, CMIP5 about 2 PB~\cite{cmip6-plan}, and CMIP6~\cite{cmip6-dashboard} exceeding 28 PB. Such an increase in volume poses unprecedented challenges to store, process, and analyze the data.

Data compression, particularly lossy compression, has emerged as a critical tool to mitigate these challenges for scientific applications~\cite{sz16, sz17, qoisz, mgard, sperr, zfp, cuZFP, zfp-particle1, zfp-particle2, tthresh}, enabling scientists to fully harness the ever-increasing performance of new computing systems. Lossy compression reduces data size by approximating original information and discarding less critical details. This process results in smaller file sizes with an acceptable loss of fidelity. Unlike domains such as natural images and videos, lossy compression designed for scientific scenarios often includes the ability to restrict the maximum point-wise error, which is essential for preserving the accuracy of scientific computations. For example, the SZ lossy compressors~\cite{interp} can reach 10$\sim$100 times higher compression ratios than lossless alternatives while keeping the maximum error within a predefined bound~\cite{sdrbench}.

\textit{\underline{Motivation:}}
Although many scientific lossy compression solutions have been proposed in the recent decade, most of them only support decompression to a single fidelity level once the data is compressed, which restricts their broad utilization.
On the one hand, scientific analyses often require different data fidelity levels as the subject of the study or the stage of investigation has diverse tolerance for data precision. For example, in hydrodynamic simulation, reconstructing viscosity may require a $2^{-5}$ finer precision compared with reconstructing vorticity from the same data field~\cite{peter-tvcg24-pframework}.
As a result, without progressive capability, researchers must compress data conservatively at the highest possible fidelity -- even though only a handful of analyses truly require it -- ultimately diminishing the overall effectiveness of data reduction.
On the other hand, when analyzing multiple snapshots, fields, or regions, researchers often first identify patterns or areas of interest at a coarse level before committing resources to detailed analysis~\cite{peter-tvcg24-pframework, pmgard, pmgard-qoi}. Without the ability of progressive retrieval, scientists have to always load the entire compressed data and decompress it at full precision. This not only increases the time and resources required for data loading and decompression but also delays subsequent analyses and scientific discoveries.

\textit{\underline{Limitation of state-of-art approaches:}}
Despite the necessity of progressive retrieval in scientific compression, as highlighted in the above scenarios, supporting this functionality is challenging due to several factors. 
First, achieving both high fidelity and high compression ratios simultaneously is challenging, as these objectives often depend on fundamentally different algorithms, where optimizing one may come at the expense of the other.
Second, straightforward progressive techniques often introduce significant operational overhead, requiring multiple passes of decompression and reprocessing -- contradicting the goal of progressive decompression, which is to save time and resources.
Third, it is already non-trivial to guarantee error bounds for partial decompression, not alone to say the progressive technique which requires the data accumulated from multiple levels to be within acceptable error margins.
Consequentially, few such progressive solutions exist, and they often fall short due to low compression ratios, high operational costs, and a lack of stringent error restrictions. These limitations hinder their adoption in scientific applications.

\textit{\underline{Key insights and contributions:}}
In this paper, we propose the first high ratio, fast, and error-bounded progressive lossy compression solution based on the interpolation algorithm. Our contributions are three-fold:
\begin{itemize}
\item We thoroughly analyze the error characteristics to build a progressive compressor based on the interpolation algorithm and propose our progressive solution, IPComp, based on the prediction model with multi-level bitplane and predictive coding. Besides having high effectiveness on retrieval, our solution supports retrieval under arbitrary error-bound settings, and it only executes decompression once for each retrieval request, compared with residual-based alternatives, which support a limited number of error bounds and require multiple passes of decompression for a single request.
\item We derive optimized strategies towards minimum data retrieval under different fidelity levels indicated by users through error bounds and bit rate. Our strategies are highly effective while being extremely lightweight with negligible overhead to the scientific workflow.
\item \textit{\underline{Experimental methodology and artifact availability:}} We evaluate the proposed solution using six real-world datasets from four diverse domains over four state-of-the-art compressors. Experimental results demonstrate our solution archives up to $487\%$ higher compression ratios and $698\%$ faster speed than other state-of-the-art progressive compressors and reduces the data volume for retrieval by up to $83\%$ compared to baselines under the same error bound and reduces the error by up to $99\%$ under the same bitrate. The source code of IPComp will be available to the public upon acceptance of the paper.
\end{itemize}

\textit{\underline{Limitations of the proposed approach:}}  
The primary focus of our solution in this paper is to design an effective progressive approach that is universally applicable across different hardware platforms. As a result, our method does not include hardware-specific optimizations yet, such as speed acceleration using tensor cores. Incorporating such optimizations will be part of our future work.

The remainder of the paper is structured as follows. In Section \ref{sec:related}, we provide an overview of related work. Section \ref{sec:overview} formulates the research problem and the overview of our design. Our developed progressive compressor is detailed in Section \ref{sec:design} to Section \ref{sec:optimizer}. Section \ref{sec:evaluation} presents and discusses the evaluation results. Finally, we draw conclusions in Section \ref{sec:conclusion}.


\section{Related Work}
\label{sec:related}
In this section, we survey existing approaches to scientific lossy compression and discuss methods that apply progressive retrieval to enhance these compression techniques.

Scientific lossy compressors aim to reduce data size while guaranteeing that the loss or error remains under a specified threshold required by the applications. 
Many scientific lossy compressors apply statistical models, including linear regression~\cite{sz17}, interpolation~\cite{interp, qoz, qoz2}, and neural networks~\cite{coordnet, aesz, srnnsz} to predict the value of the variable based on the value of the coordinates. The statistical model will be stored to reconstruct the value such that the storage of original data could be eliminated. To bound the error, the difference between prediction and real value is quantized and stored together with the model. 
On the other hand, instead of finding prediction models, some compressors choose to transform data to another domain that is more compressible, and then keep partial of the transformed data based on the error bound. Examples of transformation methods include wavelet transform~\cite{sperr}, orthogonal discrete transform~\cite{zfp}, and singular value decomposition~\cite{tthresh}. 

Progressive techniques for compression have evolved along two main directions: multi-resolution approaches, which produce output in various sizes, and multi-precision approaches, which generate output in various precisions.

For multi-resolution approaches, researchers have employed tree structures~\cite{zfp-particle2, precision-resolution-tree}, adaptive meshes~\cite{peter-mesh-tvcg22}, and wavelet transforms~\cite{li2019vapor} to partition the spatial domain hierarchically, which enables gradual refinement of data resolution. Such techniques are primarily designed for space partitioning and visualization tasks. 
Although they can support a variety of compression algorithms by applying them to each partitioned block, these methods often suffer from low compression ratios and slower speeds due to the storage and computational overhead introduced by the hierarchical structures. More importantly, they could not bound the error in each data point since the output is smaller than the original.

On the other hand, in the direction of multi-precision retrieval, one straightforward solution is to progressively refine the lossy compression error~\cite{peter-tvcg24-pframework}. 
Specifically, this involves executing compression multiple times, with each pass compressing the residual error from the previous pass but with a smaller error bound. 
This progressive strategy is orthogonal to the underlying compression method and offers versatility. However, it does not fully exploit the strengths of the base algorithm, and it incurs high operational costs as decompression must be executed multiple times to achieve a given fidelity level. As a result, researchers also aim to exploit progressive characteristics inherent in specific compression algorithms. PMGARD~\cite{pmgard, pmgard-qoi}, for instance, is a progressive solution based on the MGARD compressor. Such a solution may lead to sub-optimal performance, as shown in~\Cref{sec:evaluation}, because its underlying MGARD algorithm has lower compression ratios and speed than other state-of-the-art in many datasets~\cite{interp, qoz, qoz2}.

\section{Overview}
\label{sec:overview}
In this section, we first discuss the problems we are targeting, and then propose our solution for such problems. 

\subsection{Problem Formulation}
\label{sec: problem formulation}

\subsubsection{Definitions}
\label{sec:definitions}

Here we list five commonly adopted metrics for evaluating scientific lossy compression. The scientific dataset~\cite{sdrbench} is denoted as $x$. All symbols used in the paper are explained in~\Cref{tb:symbols}.

\noindent \textbf{Compression Ratio (CR) and Bitrate}: CR compares the original data size with compressed data size by $CR = \frac{size(\text{original data})}{size(\text{compressed data})}$. Bitrate is reverse proportional to CR. It measures the average number of bits for storing each scalar value in the compression format. A higher compression ratio or lower bit rate indicates more efficient compression in terms of storage space.
    
\noindent    \textbf{Decompression Error} describes the deviation between the original and decompressed data. There are many ways to quantify the deviation, while the $L_\infty$ norm, defined as the maximum point-wise difference between original and decompressed data, is the most widely used one. 

\noindent \textbf{Error Bound ($eb$)} is a user-defined parameter that specifies the maximum allowable error produced by lossy compressors. 

\noindent    \textbf{Peak Signal-to-Noise Ratio (PSNR)} assesses the fidelity of the data with lossy error. The definition is $20 \cdot \log_{10} \left( \frac{\max(x) - \min(x)}{\sqrt{MSE(x, \hat{x})}} \right)$, where $MSE(x, \hat x)$ denotes the mean squared error between the original dataset $x$ and the decompressed dataset $\hat x$. A higher PSNR value indicates better data fidelity.

\subsubsection{Objectives}

\begin{table}[ht]
\vspace{-4mm}
\centering
\footnotesize
\caption{Definition of symbols}
\vspace{-3mm}
\begin{adjustbox}{width=\columnwidth}
\begin{tabular}{|c|l|}
        \hline
        \textbf{Symbol} & \textbf{Description}  \\
        \hline
        $n$             & Numbers of elements of the input \\  \hline
        $x$             & The input dataset \\ \hline
        $\hat{x}$       & The decompressed output with lossy error\\ \hline
        $y,\hat y$      & The decorrelated data and its lossy version \\ \hline
        q & The quantized data \\        \hline        
        $eb$            & The bound of maximum lossy error  \\        \hline
        T, P & The transform and prediction function \\        \hline
        Q & The quantization function \\        \hline
        $\|\cdot\|_\infty$  & $L_\infty$ norm - the maximum absolute value of the input \\        \hline
        $V_l$             & The sublinear space of the l-th level. \\
                          & $V_l$ and $V_m$ are orthogonal to each other when $l \neq m$\\ \hline
       $\Pi_l$             & The $L_2$ norm projection operator of $V_l$\\ \hline
       $x_i, y_i$    & The subscript $i$ denotes the i-th component of the vector. \\ \hline
        $x_l$           & the vector is projected to $V_l$. $x_l = \Pi_l x$, which \\
                        & applies to any vector with subscription $l$\\
        \hline

\end{tabular}
\end{adjustbox}
\label{tb:symbols}
\end{table}

\begin{figure}[ht]
\vspace{-4mm}
    \centering
    \includegraphics[scale = 0.35]{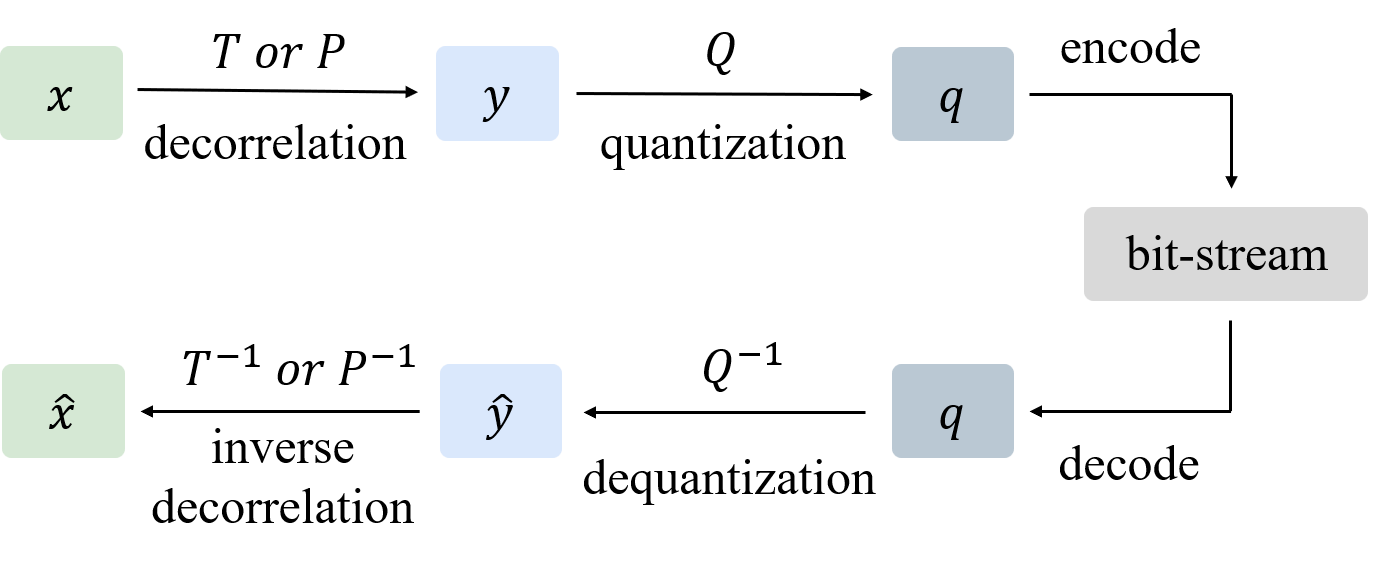}
    \setlength{\abovecaptionskip}{1mm}
    \caption{A typical lossy compression workflow. T/P represents decorrelation and Q means quantization. The quantization stage is lossy thus $\hat y$ is the lossy version of $y$. Definitions for $x$, $y$, etc can be found in~\Cref{tb:symbols}}
    \label{fig:transform} %
\end{figure}

The primary objective of this work is to develop a progressive lossy compression framework tailored for scientific data that meets the following four key goals.

\begin{itemize}
        \item \textbf{High compression ratio}: We aim to reach compression ratios higher than or equal to all other SOTAs, as a high compression ratio (CR) is essential for compressors to be effectively utilized in the scientific field. Some progressive designs, such as SZ3-M shown in~\Cref{sec:evaluation}, sacrifice compression ratio for progressiveness, but we argue that the low compression ratio will prevent the adoption of such solutions and thus make their progressive capability useless.
    \item \textbf{Progressive retrieval}: We aim to support users to retrieve the output at a lower fidelity level \( F_1 \) and incrementally refine it to higher fidelity levels \( F_2, \dots, F_n \) without the need for multiple decompression passes.
    \item \textbf{Fast speed}: Our speed of compression and decompression should be equal to or faster than all other SOTAs, and the decompression process should require only a single pass to retrieve data at any specified fidelity level \( F_i \).
    \item \textbf{Error guarantee}: We guarantee that for each fidelity level \( F_i \), if \( F_i \) has error bound restriction of \( \epsilon_i \), the reconstruction error \( E_i \) satisfies \( E_i \leq \epsilon_i \). This ensures that the compressed data remains within acceptable accuracy limits for scientific computations.
\end{itemize}

To the best of our knowledge, no compressor simultaneously achieves all these objectives yet.

\subsection{Overview of IPComp}

\begin{figure}[ht]
\vspace{-2mm}
\centering
\raisebox{-1cm}{\includegraphics[scale=0.6]{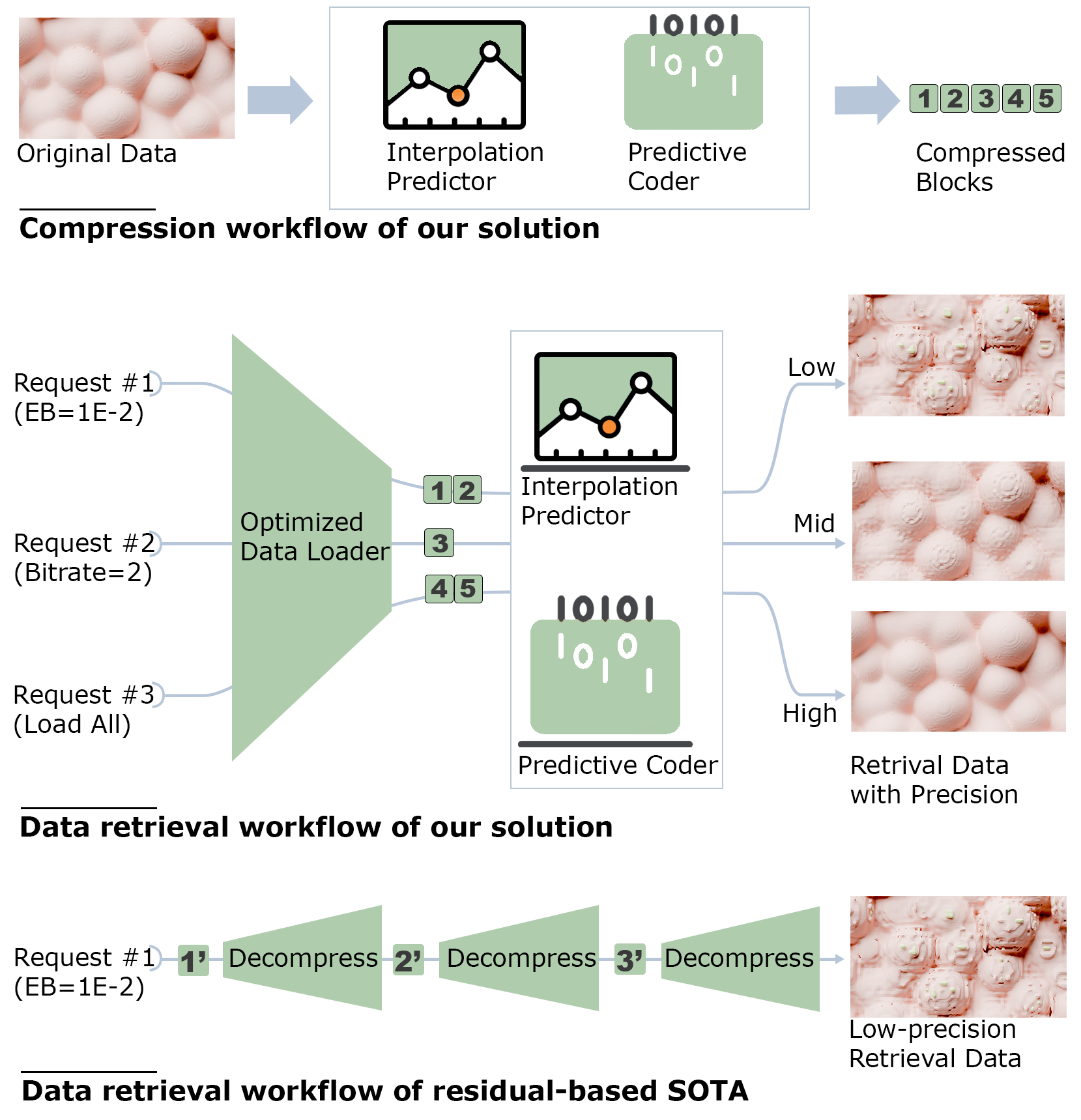}}
\caption{Overall design of our solution IPComp (the compressed data contains multiple decompressible blocks, represented as 1-5 in the diagram)}
\label{fig:overall-design}
\end{figure}

\begin{figure*}[ht]
\centering
\raisebox{-1cm}{\includegraphics[scale=0.7]{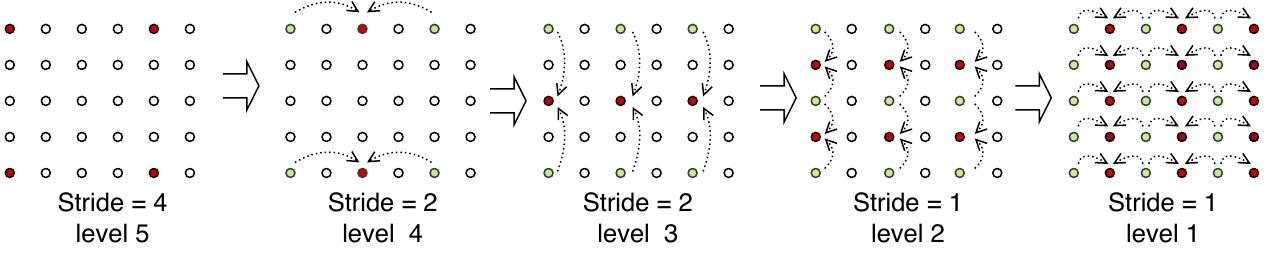}}
\vspace{-2mm}
\caption{Illustration of how the none-progressive interpolation algorithm works for a 2d input -- target points (in red color) are predicted from nearby known points (in green color), indicated by arrows}
\label{fig:interp-design}
\vspace{-2mm}
\end{figure*}

We introduce IPComp, a progressive lossy compressor designed to efficiently meet the diverse precision and fidelity requirements of scientific applications. IPComp is based on the interpolation algorithm which has been verified as the leading non-progressive reduction method, and we are the first to make it progressive. IPComp mainly has three innovative modules, shown in~\Cref{fig:overall-design}.
\begin{itemize}
    \item The \textbf{interpolation predictor} can reconstruct data to multiple fidelity levels progressively, and only needs a single pass of decompression for each retrieval request.
    \item The \textbf{predictive coder} can compress the interpolation output to independent bitplanes with high compression ratios.
    \item The \textbf{optimized data loader} can determine the minimum set of data for reconstruction to satisfy various fidelity requirements.
\end{itemize}

We demonstrate the compression and decompression workflow of IPComp in~\Cref{fig:overall-design}. For comparison, we also highlight the differences between IPComp and residual-based SOTAs.

The compression workflow begins with the original dataset, which is processed by an \textbf{Interpolation Predictor} to decorrelate the dataset based on spatial and numerical relationships, and quantize the output from floating point format to integers based on the error bound. Following this, the \textbf{Predictive coder} encodes the prediction integers by bitplanes into multiple independent blocks, each contributing incrementally to the accuracy of the decompressed data. 

The data retrieval workflow is designed to support multi-fidelity retrieval requests efficiently, leveraging an \textbf{Optimized Data Loader} that extracts only the required blocks for a given query. \Cref{fig:overall-design} demonstrates how the workflow works using three requests in increasing fidelity orders. Requests 1 asks for an error bound of 1E-2, so the data loader retrieves the first few precision blocks (e.g., blocks 1, 2) and passes them to the \textbf{Interpolation Predictor} and \textbf{Progressive coder}. These blocks are processed together at one single pass to generate low-precision output, suitable for quick exploration or coarse analyses. For the second request which indicates a targeting bitrate of 2, the data loader determines one additional block (e.g., block 3) is optimal for such a request. As a result, the predictor and coder reconstruct this mid-precision representation using block 3 on top of the low-precision result. Finally, the last request asks to retrieve all blocks, so blocks 4 and 5 are loaded consequently to build the high-precision reconstruction. 

In contrast, residual-based SOTA approaches involve significant computational overhead and redundant operations. Those approaches will first compress in a large bound, then compress the residual from the last time using a smaller bound, and repeat such residual compression until a targeted bound is reached. As a result, they require multiple decompression passes to fulfill a single retrieval request. For example, in the SOTA workflow of~\Cref{fig:overall-design}, the error bound for blocks $1'$, $2'$, $3'$ are 1, 1e-1, and 1e-2, respectively. For a single request of 1e-2, those approaches need to load three blocks and execute the decompression three times to apply the result on top of each block.

In terms of implementation, our solution is developed using the FZ framework~\cite{fz}, a premier solution for developing scientific compressors.
FZ is a comprehensive code platform providing various existing compression-related techniques wrapped in ready-to-use modules, such as the tool to compute lossy error metrics, and the function to parse compression settings from the command line and configure files. Moreover, to help developers with testing and integration, FZ provides a universal and robust API in many languages, including C, C++, Python, and Fortran, as well as seamless integration with I/O libraries like HDF5~\cite{hdf5}. 
In conclusion, FZ significantly reduces our need to re-implement existing techniques or tools, allowing us to concentrate on designing and implementing innovative progressive compression features.

\section{Progressive design of IPComp}
\label{sec:design}
In this section, we first analyze the interpolation-based algorithm, which is the leading non-progressive scientific compression strategy, and then we propose our solution to make it progressive by the prediction model and predictive coder.

\subsection{Introduction to none-progressive interpolation algorithm}
\label{sec:design intro-to-interp}

Most scientific lossy compression workflows are composed of three key steps -- decorrelation, quantization, and encoding~\cite{interp, zfp, sperr, mgard}. As shown in~\Cref{fig:transform}, the input data is defined as a vector $x$ in a linear space $\mathbb{R}^n$, where $n$ represents the total elements in the dataset. In the decorrelation step, a transform $T$ or prediction function $P$ is applied to $x$, resolving a decorrelated vector $y$. After a lossy quantization $Q$ on $y$, the quantized integers are further lossless coded to bitstream as the final compressed data. Accordingly, the decompression process executes the three steps reversely.

One critical aspect of designing an effective lossy compressor is to construct the best-suit decorrelation algorithm. Our solution IPComp uses interpolation-based decorrelation, which has been proven to be the leading solution in the scientific domain~\cite{interp, qoz, qoz2, mgard}. Its core idea is to estimate unknown data points at fixed relative indices using interpolation formulas, such as linear interpolation and cubic spline interpolation. Unlike many other predictors (such as regression-base ones) that store coefficients for reconstruction, the interpolation approach eliminates such storage overhead by relying on predefined prediction formulas for fixed indices. As a result, this approach achieves much higher compression ratios than many alternatives~\cite{interp}. 

In a plain example, consider four data points located equidistantly, with indices $i-3$, $i-1$, $i+1$, and $i+3$. Their corresponding values are denoted as $x_{i-3}$, $x_{i-1}$, $x_{i+1}$, and $x_{i+3}$. The goal is to estimate $x_i$ using interpolation formulas. 

For linear interpolation~\cite{interp}, the estimation is computed as the average of the neighboring points $x_{i-1}$ and $x_{i+1}$, given by:
\begin{equation}
y_i = \frac{1}{2}(x_{i-1} + x_{i+1}).    
\label{eq:interp-linear}
\end{equation}

For cubic spline interpolation~\cite{interp}, which takes into account all four points to achieve higher accuracy, the estimation can be:
\begin{equation}
y_i = -\frac{1}{16}x_{i-3} + \frac{9}{16}x_{i-1} + \frac{9}{16}x_{i+1} - \frac{1}{16}x_{i+3}.
\label{eq:interp-cubic}
\end{equation}

Both of the interpolation formulas predict values using neighboring data points in fixed relative positions, such that the coefficients are always the same (e.g., $\frac{1}{2}$ in the linear case) and there is no need to save them during compression.
\Cref{fig:interp-design} shows how to extend the interpolation algorithm from one scalar value to a multi-dimensional dataset. The stride separates the data points by distance (the data distance in stride i is $2^i$). In each stride, the algorithm operates recursively along each dimension of the dataset. 


\subsection{Transform vs. prediction models for progressive design}
\label{sec: design transform-vs-predict}
The interpolation-based decorrelation discussed in~\Cref{sec:design intro-to-interp} can be employed either as a transform model or a prediction model in lossy compression. In this section, we discuss our rationale for using it as a prediction model in IPComp.

Because we aim to build a progressive lossy compressor, it is crucial to account for the distortion introduced by any lossy operation, as such distortion can accumulate over successive data retrievals. Toward this, we first highlight the differences between the transform and prediction approaches, then provide a theoretical analysis of their respective distortion behaviors, and finally explain why IPComp adopts interpolation as its prediction model.


\subsubsection{Transform models}

Transform models can be viewed as a linear mapping between original data and decorrelate data. Many transform models, including Fourier transform and wavelet transform, have been adopted in lossy compression. For example, ZFP~\cite{zfp} uses nearly orthogonal block transform, and SPERR~\cite{sperr} is based on the cdf97 wavelet.

Deriving the distortion between input data $x$ and decompressed data $\hat{x}$ in such solutions is relatively straightforward. If measured by $L_\infty$ norm, the distortion can be computed as:
\[
\|x - \hat x\|_\infty = \|T^{-1}y - T^{-1}\hat y \|_\infty \leq  \|T^{-1}\|_\infty \|y - \hat y\|_\infty
\]

The value of $L_\infty(T^{-1})$ depends on the specific transformation function. Taking the basic but widely used transform function $x_i=x_i-x_{i-1}$ as an example, its corresponding transform and inverse transform formulas are:
\[
T = 
\begin{bmatrix}
1 & 0  & \cdots & 0 \\
-1 & 1 & \cdots & 0 \\
\vdots   & \vdots & \ddots & \vdots \\
0 & 0 & \cdots & 1
\end{bmatrix}, \quad
T^{-1} = 
\begin{bmatrix}
1 & 0   & \cdots & 0 \\
1 & 1   & \cdots & 0 \\
\vdots   & \vdots & \ddots & \vdots \\
1 & 1  & \cdots & 1
\end{bmatrix}
\]
In this case, the $L_\infty$ norm of $(T^{-1})$ equals the maximum sum of the rows in the matrix $T$, which equals the input size $n$.  As a result, any distortion in the transformed domain will be amplified in the original domain, and the maximum error could become $n$ times the maximum error in the transformed domain.  Given that $n$ in most scientific datasets is at least $10^7$, such a degree of distortion amplification may be unacceptable for progressive compression.

\begin{equation}
\|x - \hat x\|_\infty \leq  L_\infty(T^{-1})L_\infty(\hat{y} - y) = n \|\hat{y} - y\|_\infty
\label{eq:error-transform-lorenzo}
\end{equation}


\subsubsection{Prediction models}



Prediction models are more complex than transform models, as they tightly couple the decorrelation function $P$ with quantization $Q$. For transform models, the whole dataset is first transformed by $T$ before applying quantization $Q$. However, in prediction models, the data usually is split into orthogonal layers or groups, and the prediction $P$ and quantization $Q$ are applied layer by layer~\cite{di2024survey, mgard} on $\hat{x}$ (data with lossy error) instead of $x$ (original data). As a result, the prediction model can be viewed as a non-linear mapping between input and decorrelate data.

\Cref{fig:interp-design} shows how to divide the whole linear space into multiple orthogonal levels by setting a shrinking stride distance. We represent level as the sublinear spaces $V_l \subseteq \mathbb{R}^n$. The total number of levels is defined as $L$.
The prediction $
P_l
$ will be executed $L-1$ times, covering from the top level $L$ to level 1.


For each level $l$, we process $y_l$ which is the difference between the original data $x_l$ and its prediction. The prediction $P_l$ is based on interpolation from $\hat{x}_{l+1}$ (the previous level's data with lossy error).
\[
y_l = x_l - P_l\hat{x}_{l+1}
\]
The prediction difference $y_l$ will be quantized to integers, and dequantized back to floating point $\hat y_l$. The quantization guarantees that the point-wise error between $y$ and $\hat y_l$ is less than the pre-defined global error bound $eb$. Note that in decompression we do not have the original prediction difference $y$ since it is lossy quantized here.
\[
y_l - \hat y_l = \varepsilon_l,\|\varepsilon_l\|_\infty \leq eb
\]
In decompression, which is the reverse process, we first recover the prediction data $P_l\hat{x}_{l+1}$, then compute the output $\hat{x}$ by adding $\hat y$ (the prediction difference with error) on top of the prediction. 
\[
\hat{x}_l = P_l\hat{x}_{l+1} + \hat{y}_l
\]
By simplifying the three formulas above, we can get the conclusion that for each level, $\hat x_l$ only differs with original data $x_l$ within the point-wise error bound $eb$.
\[
x_l = \hat{x}_l + \varepsilon_l, \|\varepsilon_l\|_\infty \leq eb
\]
Then for the whole dataset, the maximum point-wise error could be bounded by 
\begin{equation}
    \|x-\hat x\|_\infty = \max_l\|x_l -\hat x_l\|_\infty = \max_l \varepsilon_l \leq eb
\label{eq:error-prediction-interp}
\end{equation}

Compared with this~\Cref{eq:error-prediction-interp}, the error in transform models (shown in~\Cref{eq:error-transform-lorenzo}) is proportional to the data size, which means it will be very difficult to bound the error in transform models if the input data is large. Since progressive may cause the error to accumulate across retrievals, we want to have an error control as tight as possible. As a result, we choose to use interpolation as a prediction model for our progressive solution.


\subsection{Progressive interpolation algorithm}
\label{sec: design progressive}

\begin{figure}[ht]
    \centering
    \hspace{-6mm}
    \includegraphics[scale = 0.5]{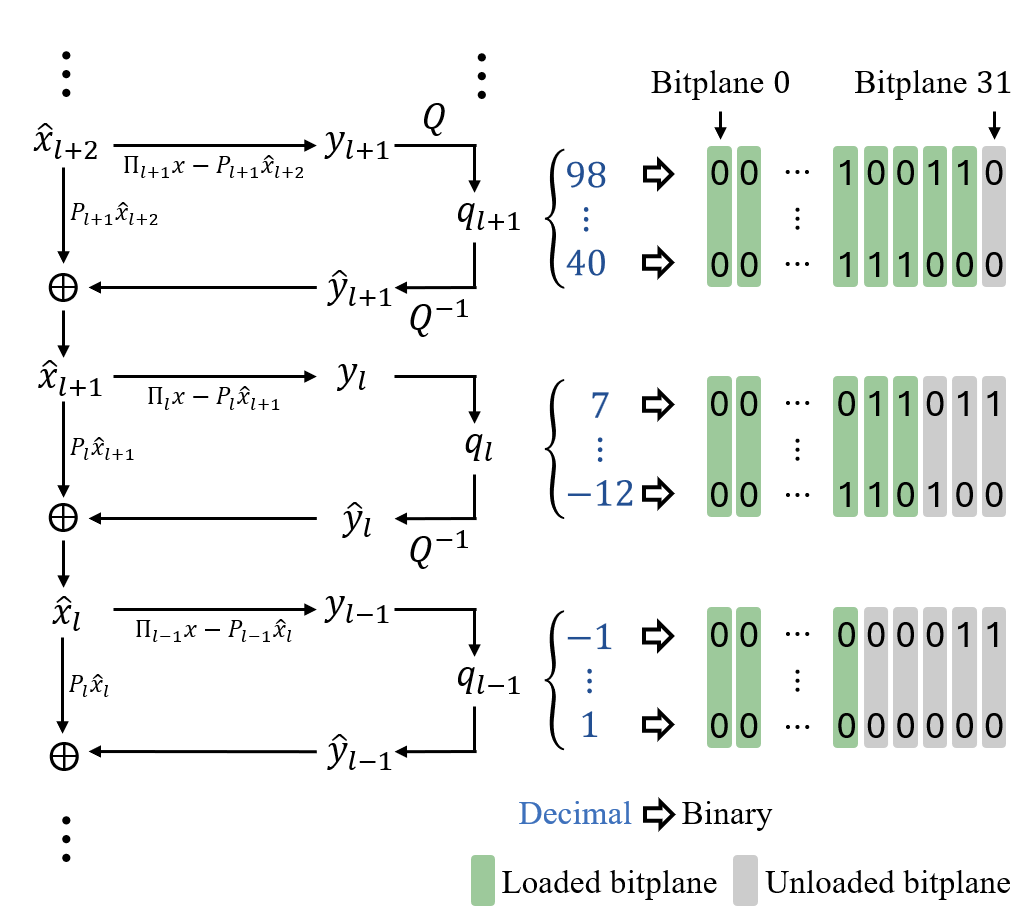} 
    \vspace{-5mm}
    \caption{Our progressive solution splits the quantization integers by bitplanes and encodes them separately} %
    \label{fig:overwrite} %
\end{figure}

In this section, we present our solution to support progressive retrieval utilizing the interpolation-based prediction model discussed in~\Cref{sec:design intro-to-interp} and~\Cref{sec: design transform-vs-predict}.

The foundation of our progressive design is to load a partial of the information of $\hat y$ according to the bitplane. As illustrated in~\Cref{fig:overwrite}, in our approach, the quantized data (denoted as $q_l$ at level $l$) are 32-bit integers. The bits from the same position across multiple quantized integers form a bitplane (represented by small rectangular boxes in the figure). We independently encode the 32 bitplanes at the same level, which allows lower-fidelity outputs to be reconstructed by loading only some of the bits rather than all of them.
When retrieving data, our optimizer discussed in~\Cref{sec:optimizer} will determine the optimal loading strategy -- the minimum bitplanes to load from each level to meet the retrieval requirement. 

\Cref{alg:reconstruction} describes the reconstruction process of our solution. 
As shown in~\Cref{fig:overwrite}, the reconstruction starts from the top level, where the loaded data is decoded into the reconstructed diff $\hat{y}$.
We then use the $\hat{y}$ from each level, in sequential order, to progressively reconstruct the original data. 
The parameter $L_p$ indicating in which level we start progressive compression. The first step from line 3 to 7 is to retrieve data of non-progressive levels. Then the second part is to load the bitplanes we are requiring, and decoded them in addition to the vector we get in the first step. 

~\Cref{alg:reconstruction} is designed for cases where data is reconstructed from scratch. On the other hand, when users find the current precision is insufficient, \Cref{alg:incremental} shows how to update the output from lower precision to higher precision by loading incremental bitplanes. 
By combining these newly loaded bitplanes with the previously loaded data, our solution can reconstruct a higher fidelity output without reloading all the compressed data.

\begin{algorithm}
\caption{Reconstruction Algorithm}
\label{alg:reconstruction}
\begin{algorithmic}[1] 
  \Require $bitplaneList[L]$
  \State $\hat x \gets 0$
  \State $\Pi_L \hat x \gets P_L(\mathbf{0})$
  \For{$l \gets L - 1$ \textbf{downto} $L_p + 1$}
      \State $q_l \gets \textbf{decode}(bitplaneList[l])$
      \State $\hat{y}_l \gets \textbf{dequantizition}(q_l)$
      \State $\hat x_l \gets \textbf{Predict}(\hat{x}, \hat y_l)$
  \EndFor
  \State $\Delta \gets 0$
  \For{$l \gets L_p$ \textbf{downto} $1$}
      \State $q_l \gets \textbf{decode}(bitplaneList[l])$
      \State $\hat{y}_l \gets \textbf{dequantizition}(q_l)$
      \State $\hat x_l \gets \textbf{Predict}(\Delta, \hat y_l)$
      \State $\Delta_l \gets \textbf{Predict}(\Delta, \hat y_l)$
      
  \EndFor \\
  \Return $\hat x$
\end{algorithmic}
\end{algorithm}

\begin{algorithm}
\caption{Incremental Reconstruction Algorithm}
\label{alg:incremental}
\begin{algorithmic}[1]  
  \Require $\hat x^{\text{(old)}}, bitplaneList[L]$
  
  \State $\hat x^{\text{(new)}} \gets \hat x^{\text{(old)}}$
  \State $\Delta \gets 0$
  \For{$l = L_p$ \textbf{to} 1}
    \State $q_l \gets \textbf{decode}(bitplaneList[l])$
    \State $\hat y_l = \textbf{dequantization}(q_l)$
    \State $x_l^{\text{(new)}} \gets \textbf{Predict}(\Delta, \hat y_l)$
    \State $\Delta_l \gets \textbf{Predict}(\Delta, \hat y_l)$
  \EndFor\\
  \Return $\hat x^{\text{(new)}}$
\end{algorithmic}
\end{algorithm}




\subsection{Predictive negabinary Coding}
\label{sec:design coding}

In this section, we propose a novel coding method that addresses two key challenges in progressive coding: preserving data correlation across bitplanes and handling sign bits. Our solution achieves high compression ratios by capturing cross-bitplane correlations using predictive coding and encoding sign bits using negabinary coding.

\subsubsection{Predictive Bitplane Coding}

We propose a predict-based strategy to exploit the correlation between bitgroups. To support progressive retrieval, each quantized integer is split into bitplanes, which means that the bits of a single integer are encoded separately. As a result, the correlation between bits from the same integer is totally ignored. 
We observe that during decompression when retrieving a certain bitplane at a specific level, the previously loaded bitplanes are already known. This observation allows us to leverage the correlation between bitplanes by predicting the bit value based on previous bits from the same integer, and encode the results of the prediction instead of raw bits. For example, if the prefix bit is 0, we predict that the current bit $b$ will also be 0. This prediction mechanism is equivalent to applying an XOR operation between the prefix bit and $b$ as $\text{encoded bit} = \text{prefix bit} \oplus b $.

\begin{table}[ht]
\centering
\footnotesize
  \caption{Our predictive bitplane coding strategy reduces the entropy in quantized integers (lower entropy indicates better compressibility)} 
  \vspace{-2mm}
  \label{tb:entropy} 
  \begin{adjustbox}{width=\columnwidth}
    \begin{tabular}{|c|c|c|c|c|}
        \hline
        Fields & Original & 1-bit prefix & 2-bits prefix & 3-bits prefix  \\ 
        \hline
        Density    & 0.358505 & 0.352111 & 0.34732 & 0.350525   \\
        SpeedX    & 0.868908 & 0.861458 & 0.855473 & 0.859443     \\
        Wave    & 0.440292 & 0.433391 & 0.427993 & 0.431599   \\
        \hline
    \end{tabular} 
\end{adjustbox}
\end{table}


We can further extend this method by utilizing more prefix bits -- we perform an XOR operation on all the preceding bits and then XOR the result with the target bit $b$ to obtain the encoded bit. The process can be mathematically expressed as:
\[
\text{encoded\_bit} = (\text{prefix\_bit}_1 \oplus \text{prefix\_bit}_2 \oplus ...\oplus \text{prefix\_bit}_n) \oplus b
\]

We observe from~\Cref{tb:entropy} that using 1-bit, 2-bit, or 3-bit prefix bits for prediction consistently reduces the entropy compared to the original data. Among these, 2-bit prediction generally achieves the best entropy drop. As a result, our approach leverages the two prefix bits to predict the next bit.

\subsubsection{Negabinary coding}
Unlike their non-progressive counterparts, progressive compression schemes must efficiently handle sign bits, since loading them is a prerequisite for reconstructing values, but storing sign bits separately and loading them first before all other bits incurs significant overhead. To address this issue, we evaluate the suitability of three widely used methods for encoding sign bits in a progressive setting -- two's complement~\cite{computer_organization_design}, sign-magnitude~\cite{computer_organization_design}, and negabinary~\cite{hoang2018study, zfp}, and select negabinary encoding as the preferred approach for two reasons.

First, negabinary encoding may lead to better compressibility than other methods for values that fluctuate around zero (which is predominant in the last layer of quantized integers). In two's complement, the higher-order bits often contain many 1s for values near zero, making the corresponding high-order bitplanes difficult to compress. For example, the 8-bit representations of 1 and -1 are 00000001 and 11111111 in two's complement, 00000001 and 10000001 in sign-magnitude, and  00000001 and 00000011 in negabinary. As a result, negabinary encoding keeps the higher-order bits as 0, which leads to more compressible high-order bitplanes compared to the dense 1s in two's complement.
Second, compared to sign-magnitude encoding, negabinary encoding is more balanced. When the least significant bits are set to zero (due to quantization or truncation), the resulting error uncertainty is smaller in negabinary encoding. This leads to more predictable error behavior, which is beneficial for data compression and reconstruction.
The uncertainty could be formulated by
    \[
    \begin{aligned}
       & \text{uncertainty (negabinary) } = 
        \begin{cases}
                \frac{2}{3}\times2^d-\frac{1}{3}
            , & \text{when } d \text{ is odd} \\
                \frac{2}{3}\times2^d-\frac{2}{3}, & \text{when } d \text{ is even}
        \end{cases}\\
    & \text{uncertainty (sign-magnitude) }  = 2^d-1
    \label{eq:uncertainty}
    \end{aligned}\\
    \]
where d represents the number of bits discarded. As d increases, the uncertainty of negabinary is only around two-thirds of sign-magnitude encoding.


\section{Optimized data loading} 
\label{sec:optimizer}
In this section, we present our optimized data load strategy that can minimize the volume of data loaded under the retrieval constraints.
Our optimizer supports both error-bound mode and bitrate mode which represents most of the scenarios in scientific data retrieval.
\begin{itemize}
    \item Error bound mode: in this mode, users specify the required precision and the optimizer selects the optimal strategy to ensure that the reconstructed data's error remains strictly within the given bound while minimizing the amount of data loaded.
    
    \item Fixed rate/size mode: in this mode, the user specifies a maximum allowable bitrate or size, typically constrained by I/O bandwidth. The optimizer then ensures that the reconstructed data has the smallest possible error while staying within the specified bitrate limit. Since the total data size $n$ is constant for the input, requiring a specific bitrate is equivalent to requiring a fixed retrieval size.
\end{itemize}
We first discuss the upper bound theory which both modes rely on, then show how the optimizer makes the best decisions on bitplane selection based on dynamic programming.

\subsection{Fundamental theory for our optimizer}
\label{sec: optimized-base-theory}
In this section, we discuss the fundamental theory for our optimizer which measures the error caused by only a partial of bitplanes being loaded.

\begin{theorem}
The $L_\infty$ error in progressive retrieval can be bounded based on the information loss due to the unloaded bitplanes.

\begin{equation}
\|x-\hat x\|_\infty \leq \sum_{l = 0}^{L-1} p^{l}\|\delta y_{l+1}\|_\infty + eb
\label{eq:upperbound_1}
\end{equation}

In this equation, $p=1$ for linear interpolation, and $p=1.25$ for cubic interpolation. $\delta y_l$ describes the information loss in level $l$ caused by the unloaded bitplanes, and its value can be pre-computed during compression.

\end{theorem}

\renewcommand{\qedsymbol}{}
\begin{proof}
As illustrated in \Cref{alg:reconstruction}, the retrieval process proceeds iteratively through all levels, with each level relying on the preceding level's output to inform its predictions. This introduces a significant hurdle for error analysis, as uncertainty can propagate from one level to the next. To address this, we conceptualize the error in level $i$ as being comprised of three components. The first component, $\delta y_{l}$, represents the information loss resulting from not loading certain bitplanes at this level. The second component, $\delta y_{prop, l}$, accounts for information loss carried over from the prior level. The third component, $\varepsilon_l$, captures the error introduced by lossy quantization.\begin{equation}
    x_l - \hat x_l = \delta y_l + \delta y_{prop, l} +  \varepsilon_l
\end{equation}

The information loss propagated until the previous level $l+1$ is $\delta y_{prop, l+1}$. It will be first combined with the information loss in the current level $\delta y_{l+1}$, and then transfered by prediction and passed to the next level as $\delta y_{prop, l}$.
\begin{equation}
    \delta y_{prop, l} = P_l(\delta y_{prop, l+1} + \delta y_{l+1})
\end{equation}

Combine the two equations, we can have: 
\begin{equation}
x_l - \hat x_l = \delta y_l + \sum_{m = 0} P_{l}P_{l+1}...P_{l+m} \delta y_{l+m+1} + \varepsilon_l
\label{eq:upperbound_i}
\end{equation}

In our approach, the $L_\infty$ of interpolation $P_l$ is greater or equal to 1, which implies the information loss would be amplified to the lower level. In other words, lower levels have larger errors, and the maximum error of the dataset is in the first level. As a result, we can quantify the error of the dataset with \Cref{eq:upperbound_i} by:

\begin{align}
\|x-\hat x\|_\infty &= \|x-\hat x_1\|_\infty  \notag \\
&= \|\delta y_1 + \sum_{l = 0} P_{1}P_{1+1}...P_{l} \delta y_{l+1} + \varepsilon_1\|_\infty  \notag \\
&\leq \sum\|P_1\|_\infty\|P_2\|_\infty...\|P_l\|_\infty\|\delta y_{l+1}\|_\infty + eb
\label{eq:upperbound}
\end{align}

Now we simplify \Cref{eq:upperbound} to \Cref{eq:upperbound_1}. The value of $L_\infty(P)$ depends on the prediction method. Liner interpolation, as shown in~\Cref{eq:interp-linear}, has two coefficients both at 0.5. Such that $L_\infty (P) = 0.5+0.5=1$. Similarly, cubic interpolation, as shown in~\Cref{eq:interp-cubic}, has four coefficients, and $L_\infty(P) = 2\times \frac{1}{16} + 2\times \frac{9}{16} =1.25$. With the derived value of $L_\infty(P)$, \Cref{eq:upperbound} is converted to~\Cref{eq:upperbound_1}.

\end{proof}

\subsection{Optimized loading based on error-bounds}
\label{sec: optimized-eb}

The goal of error-bound mode is to generate a loading strategy specifying the minimum number of bitplanes to load in each layer, in order to let all point-wise lossy errors be less or equal to the user-defined bound (denoted as E). Finding such bitplanes can be formalized as an optimization problem.
\begin{gather*}
\max_{b_l, l \in \{1, 2, ... L\}}  \sum_l SavedSize(l, b_l), \\
\text{subject to}   \sum_l err(l,b_l) + eb \leq \text{E}.
\end{gather*}
$b_l$ indicates the number of bitplanes discarded at the level $l$. \linebreak $SavedSize(l,b_l)$ denotes the amount of data saved by discarding $b_l$ bitplanes at level $l$. $err(l,b_l)$ denotes the error at level $l$ by discarding $b_l$ bitplanes. 
As stated in \Cref{eq:uncertainty}, $err(l,b_l) = p^{l - 1}\|\delta y_{l}\|_\infty$, where $p = 1$ for linear interpolation, $p=1.25$ for cubic interpolation, and $\delta y_l$ is a function of $b_l$.

By such definitions, we essentially reformulate this optimization problem as the classical knapsack problem, enabling it to be efficiently solved using dynamic programming (DP) with minimal computational overhead.
Let $DP(l,e)$ be the maximum total saved size when the last layer is $l$ and the maximum error is $e$. Then $DP(L, E)$ would be our best solution to load data for the given error bound E, and its value can be derived recursively by the following DP transition function:
\[
DP(l,e) = \max_{b_l, s.t. err(l,b_l) \leq e }\{DP(l-1, e - err(l,b_l)) + SavedSize(l,b_l) \}
\]
The time complexity of such a DP process is $O( \text{\#level} \times \text{\#bitplane} \times \text{\#discrete error values})$. 
The number of levels is $\log_2 n$ where n is the input size, and the number of bitplanes is 32. The discrete error values fall within the range of $[128, 1023]$ by normalized retrieval bound $E$ by compression bound $eb$. If comparing the time of the DP to the time of simply traversing the input, the ratio would be 
$\frac{(1023-128+1)*32*\log_2 n}{n} \approx \frac{3*10^4\log_2 n}{n} \approx 3\%$ for datasets listed in~\Cref{tab:apps}. Since compression takes much more computation than just traversing the input, the overhead of the DP relative to compression is totally negligible.

\subsection{Optimized loading based on bitrates}

In the fixed bitrate mode, the user specifies a maximum allowed retrieval size \(S\) (or equivalently, a maximum allowable bitrate). The optimizer then seeks to minimize the reconstruction error while ensuring that the loaded data does not exceed \(S\). 



With $LoadedSize(l,b_l)$ representing the loaded size of level $l$, this problem can be formalized as follows:
\[
\begin{aligned}
\min_{b_l, l \in \{1, 2, ... L\}} \quad & \sum_l err(l,b_l) + eb, \\
\text{subject to} \quad & \sum_l LoadedSize(l,b_l) \leq S.
\end{aligned}
\]

Similar with~\Cref{sec: optimized-eb}, we effectively convert this problem to the knapsack dynamic programming problem, so that it can be solved similarly as in the error bound mode, also with negligible overhead.

\section{Experimental Evaluation}
\label{sec:evaluation}
In this section, we describe the experimental setup and evaluate our solution on six datasets against four state-of-the-art baseline compressors.

\subsection{Experimental Setting}

\subsubsection{Execution Environment} The experiments are performed on the Purdue Anvil supercomputer~\cite{anvil} through NSF ACCESS~\cite{access}. Each computing node in Anvil features two AMD EPYC 7763 CPUs with 64 cores at a 2.45GHz clock rate and 256 GB DDR4-3200 RAM. 

\begin{table}[ht]

    \centering
    \caption{Data in our experiments}
    \vspace{-3mm}
    \begin{adjustbox}{width=\columnwidth}
    \begin{tabular}{|c|c|c|c|c|}
    \hline
Name & Explanation & Precision & Shape  \\ \hline
Density~\cite{sdrbench} & mass per unit volume in turbulence & 64 & $256\times384\times384$ \\ \hline
Pressure~\cite{sdrbench} & thermodynamic pressure in turbulence & 64 & $256\times384\times384$ \\ \hline
VelocityX~\cite{sdrbench} & x-direction velocity in turbulence & 64 & $256\times384\times384$ \\ \hline
Wave~\cite{interp} & wavefield evolution in seismic & 64 & $1008\times1008\times352$ \\ \hline
SpeedX~\cite{sdrbench} & x-direction wind speed in weather  & 64 & $100\times500\times500$ \\ \hline
CH4~\cite{sdrbench} & mass fraction of CH4  in combustion & 64 & $500\times500\times500$ \\ \hline
\end{tabular}
\end{adjustbox}
\label{tab:apps}
\vspace{-3mm}

\end{table}

\subsubsection{Datasets} The experiments are evaluated on six datasets across four diverse scientific domains, as listed in~\Cref{tab:apps}.

\subsubsection{State-of-the-Art lossy compressors in our evaluation}
\label{sec: exp-baseline}
The experiments include four SOTA scientific lossy compressors as baselines -- SZ3-M, SZ3-R, ZFP-R, and PMGARD. 

\noindent\textbf{SZ3-M}~\cite{pmgard-qoi}: SZ3-M (where "M" stands for multi-fidelity) is the straightforward multi fidelity version of SZ3 based on multiple outputs. SZ3 is the leading non-progressive scientific lossy compressor. SZ3 uses interpolation as prediction, together with linear-scale quantization, Huffman coding, and zstd lossless coding~\cite{rfc8478}. SZ3 has excellent compression ratio and fidelity over others while its speed may not be as fast as ZFP.  SZ3-M compresses the input with different error bounds independently and groups those compressed data all together as output. Such a solution supports multi-fidelity retrieval but is not progressive yet, as it cannot reuse the low-fidelity data to build high-fidelity results. 

\noindent\textbf{SZ3-R}~\cite{pmgard-qoi, peter-tvcg24-pframework}: SZ3-R (where "R" stands for residual) is the progressive version of SZ3 based on residuals. It first compresses the input with a large bound, then compresses the lossy error (residual) from the last time using a smaller bound, and repeats such residual compression until a targeted bound is reached. 

\noindent\textbf{ZFP-R}~\cite{peter-tvcg24-pframework}: ZFP-R is the progressive version of ZFP based on residuals. It handles residuals the same way as SZ3-R. ZFP~\cite{zfp} is the leading transform-based lossy compressor based on orthogonal transformation. ZFP is usually the fastest scientific lossy compressor because of its highly efficient transform function, although its compression ratio may not be as high as others.

\noindent\textbf{PMGARD}~\cite{pmgard, pmgard-qoi}: PMGARD (where "P" stands for progressive) is the progressive version of the MGARD~\cite{pmgard} compressor. MGARD~\cite{mgard, mgardx} is a multigrid-based lossy compressor that uses hierarchical decomposition to remove redundancy in scientific data while preserving error bounds.

We note that the residual-based approaches (SZ3-R and ZFP-R) have drawbacks that affect our evaluation. 
First, they have limited error-bound flexibility. More specifically, the retrieval is only possible at a few predefined error bounds. This creates a trade-off -- setting too many error bounds reduces the overall compression ratio and significantly degrades compression/decompression throughput, while setting too few nodes limits flexibility in selecting error bounds during decompression. For the experiments, we configure five error bounds for them, where each successive one is a factor of $2^2$ (4×) apart. Specifically, the error bounds are set to $2^{16}eb$, $2^{14}eb$, $2^{12}eb$, $2^{10}eb$, $2^{8}eb$, $2^{6}eb$, $2^{4}eb$, $2^{2}eb$, and $eb$.
Second, those methods are not well-suited for retrievals based on pre-defined bitrates, as the residuals are compressed by error bounds, and their size is not aligned with the bitrate of the retrieval requests. As a result, we select the largest residual that fits within the user-specified bitrate constraint for SZ3-R and ZFP-R in the experiments.

\subsection{Evaluation Results and Analysis}
The evaluation covers three aspects -- compression ratio, progressive retrieval effectiveness under error-bound or bitrate constraints, and compression and decompression/retrieval speed.  To be more specific, the compression ratio comparison demonstrates our solution IPComp leads to the smallest compressed data size. The progressive retrieval evaluation shows that IPComp requires the lowest data volume to reconstruct to the same fidelity compared with others, and IPComp leads to the highest reconstruction fidelity under the same bit rate budget. The speed test confirms IPComp's high compression efficiency, especially compared to residual-based alternatives (SZ3-R and ZFP-R) whose speed drops significantly when the number of residual compressions increases.

\begin{figure}[ht] \centering
\includegraphics[scale=0.14]{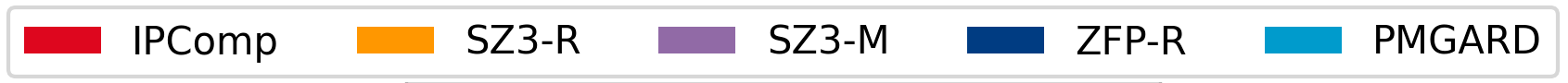}

\hspace{-8mm}
\subfigure[High precision setting ($eb = 1\mathrm{e}{-9}$)]{
\includegraphics[scale=0.48]{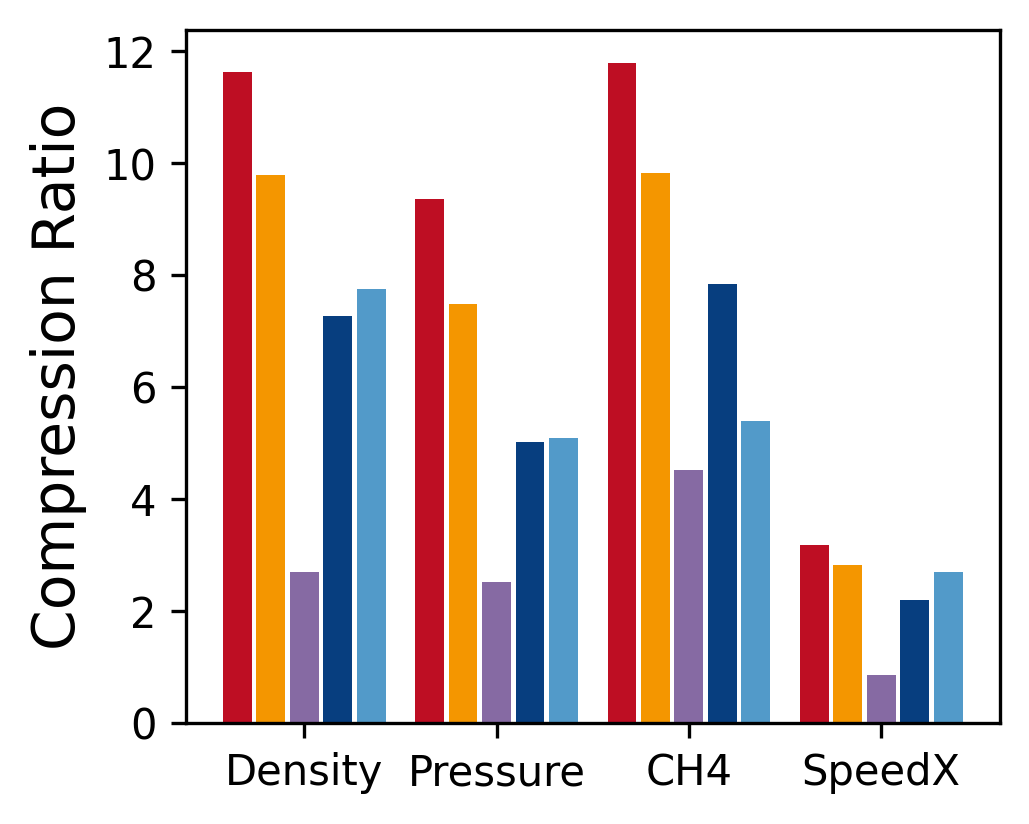}
}
\hspace{-4mm}
\subfigure[High ratio setting ($eb=1\mathrm{e}{-6}$)]{
\includegraphics[scale=0.48]{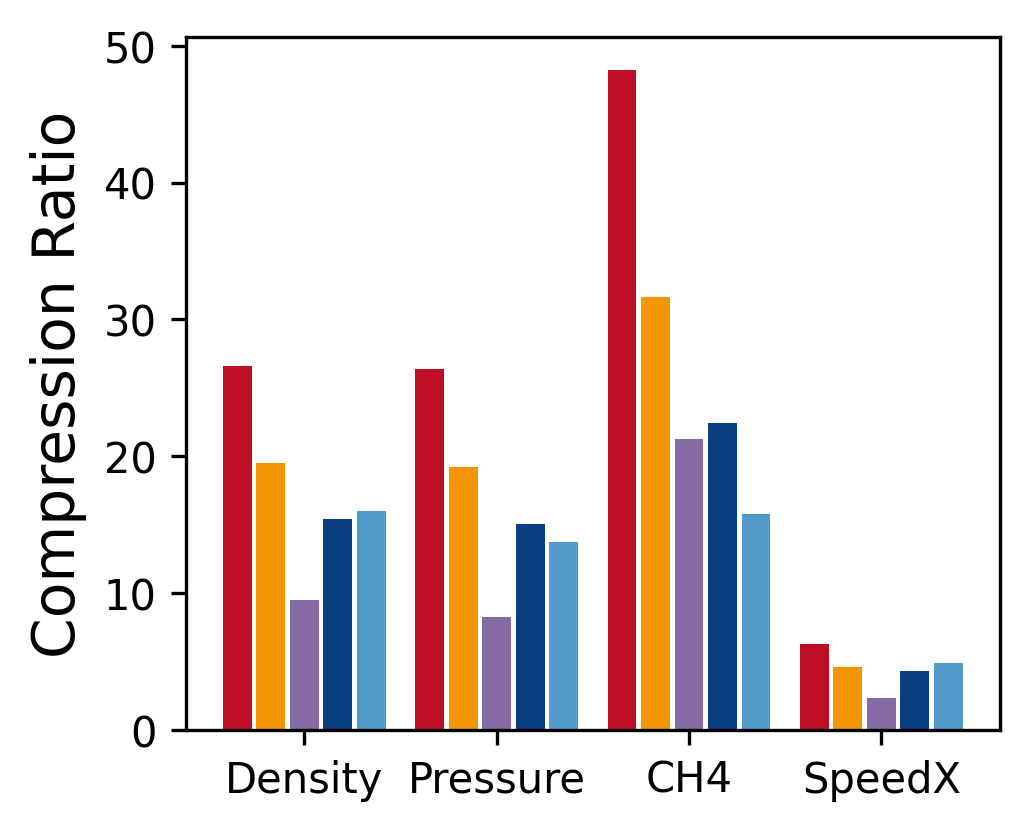}
}
\hspace{-8mm}
\caption{Our compressor IPComp leads the compression ratio among all baselines}
\label{fig:cr}

\end{figure}

\begin{figure}[ht] \centering

\includegraphics[scale=0.14]{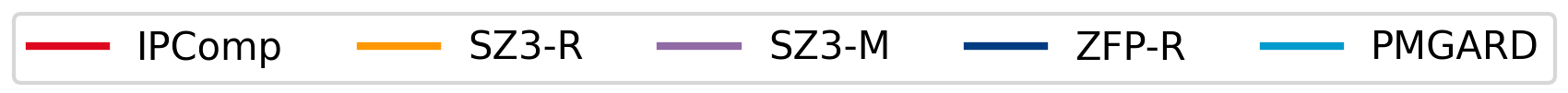}

\subfigure[Density]
{
\raisebox{-1cm}{\includegraphics[scale=0.6]{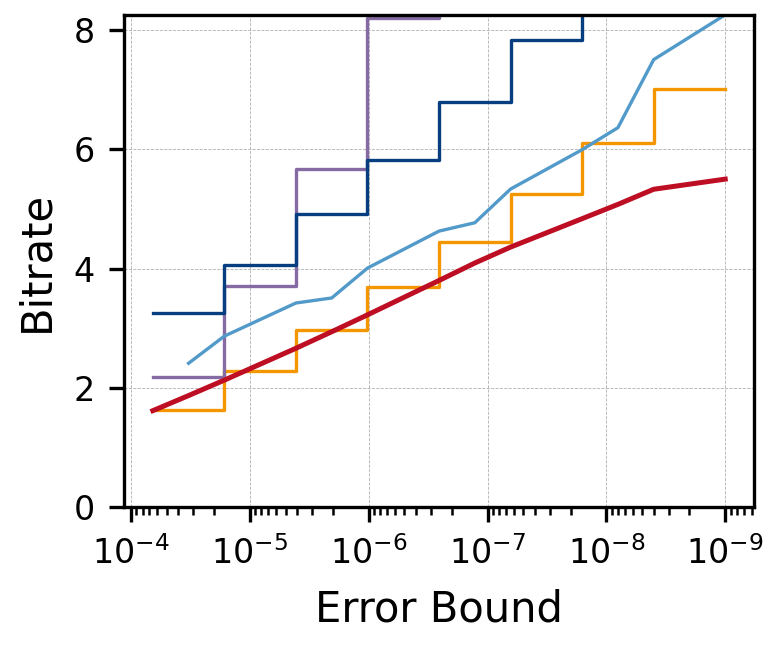}}
}
\subfigure[Pressure]
{
\raisebox{-1cm}{\includegraphics[scale=0.6]{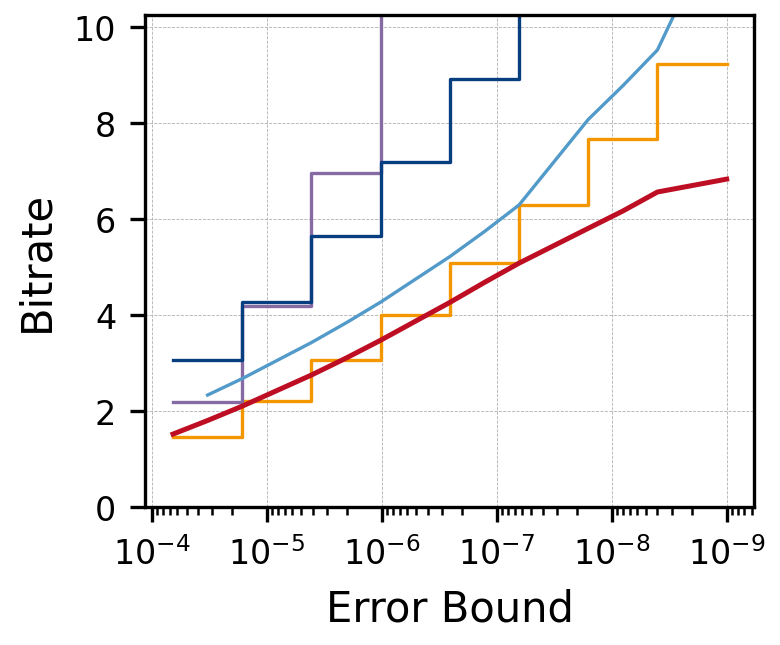}}
}
\vspace{-3mm}

\subfigure[VelocityX]
{
\raisebox{-1cm}{\includegraphics[scale=0.6]{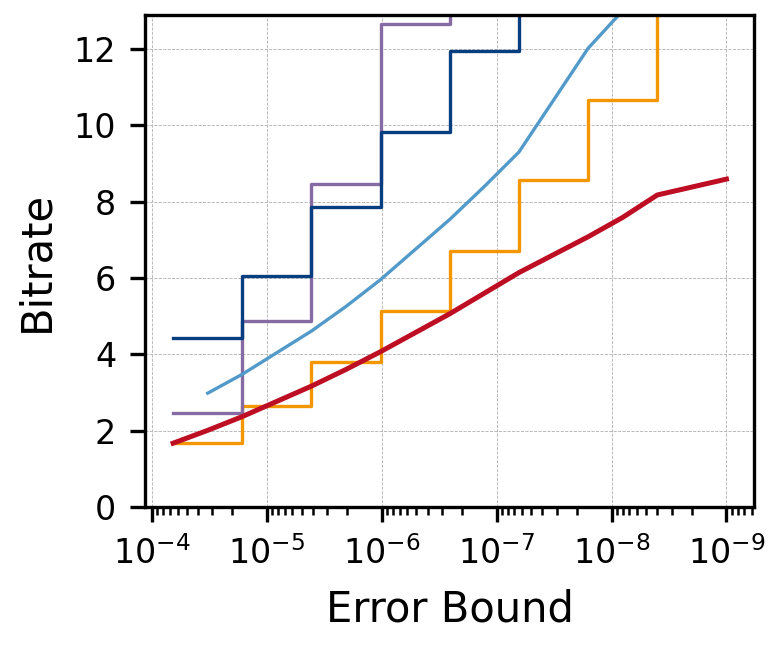}}
}
\subfigure[Wave]
{
\raisebox{-1cm}{\includegraphics[scale=0.6]{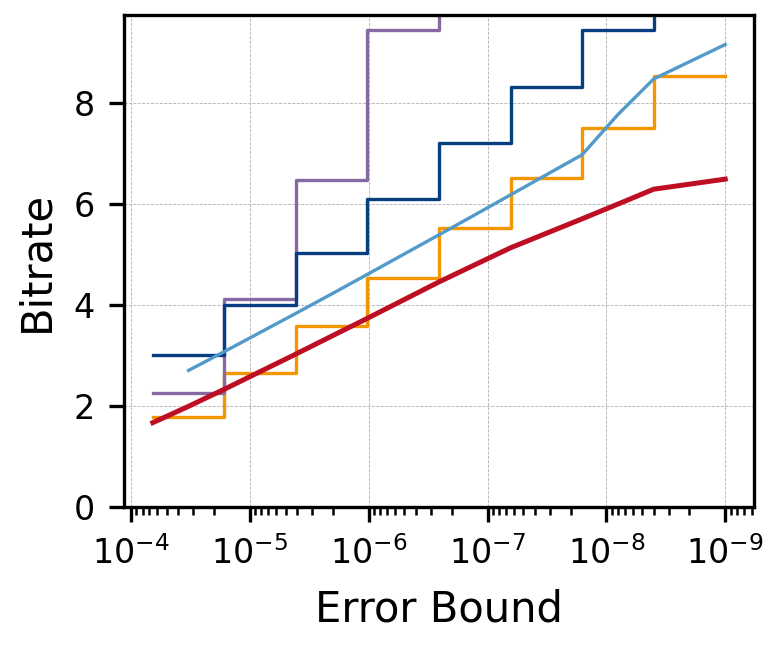}}
}

\vspace{-3mm}
\subfigure[SpeedX]
{
\raisebox{-1cm}{\includegraphics[scale=0.6]{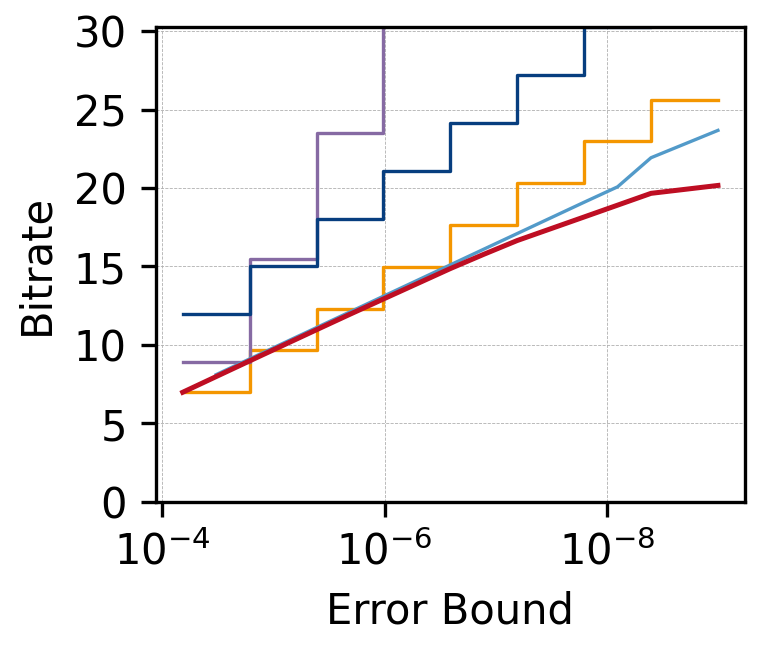}}
}
\subfigure[CH4]
{
\raisebox{-1cm}{\includegraphics[scale=0.6]{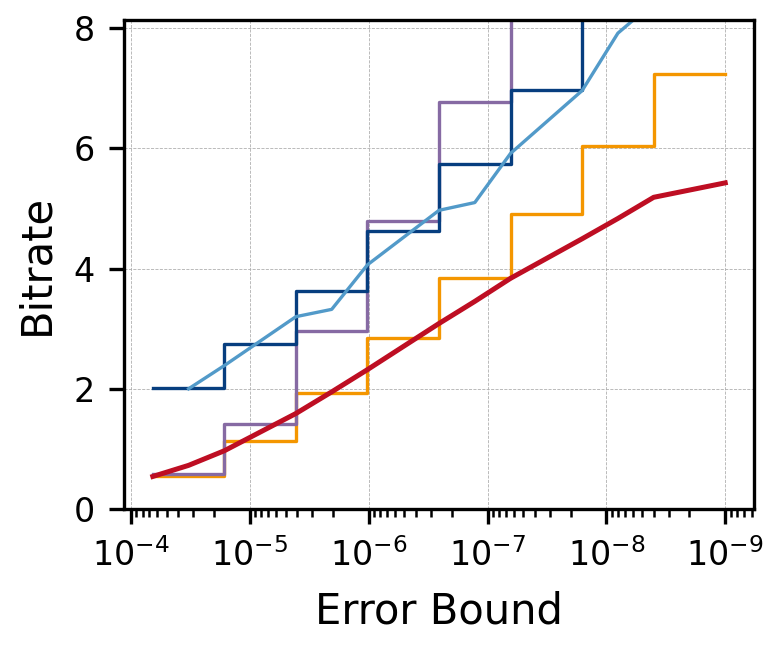}}
}
\vspace{-3mm}
\caption{Our compressor IPComp takes the smallest amount of data to reconstruct toward the same $L_\infty$ error compared with all the baselines. Moreover, IPComp supports arbitrary error bound, while SZ3-R and ZFP-R are limited to a few pre-defined bounds}
\label{fig:bitrate_eb}
\vspace{-2mm}

\end{figure}

\subsubsection{Compression Ratios}

CR is a key metric for evaluating any compressor, regardless of whether it is progressive or not. As discussed in~\Cref{sec:intro}, the fundamental goal of compression is to reduce data size, as smaller data sizes alleviate the burden of storage, transfer, and processing on applications.

\Cref{fig:cr} presents the compression ratios of all solutions under the same error-bound settings. We select two error bounds which are 1e-6 and 1e-9  to cover both high precision (cr<10) and high ratio (cr up to 50) cases.
As this figure shows, our solution IPComp achieves a compression ratio advantage of around 20\% to 500\% over other progressive compressors across the vast majority of datasets. This allows users to perform progressive decompression while utilizing the least possible storage space.


In fact, although not shown in the figure, IPComp achieves an even higher compression ratio than the non-progressive SZ3 particularly in high-precision scenarios, despite both being based on the interpolation prediction algorithm.
The primary reason for this improvement lies in the encoding process. The non-progressive SZ3 uses Huffman entropy encoding, which is then further compressed using zstd. Since Huffman coding assigns variable-length codes based on frequency, it may disrupt certain repetitive patterns in the byte or word level after encoding. This, in turn, reduces the effectiveness of zstd’s lossless compression, as zstd relies on detecting and exploiting repetitive patterns at the byte/word level to achieve higher compression efficiency. ~\cite{rfc8478}
In contrast, our solution IPComp employs a customized predictive encoding method, discussed in~\Cref{sec:design coding}, followed by zstd for entropy encoding and pattern extraction. By avoiding the disruption caused by Huffman and preserving more repetitive structures, our method enhances the final compression ratio.

\begin{figure}[ht] \centering

\includegraphics[scale=0.14]{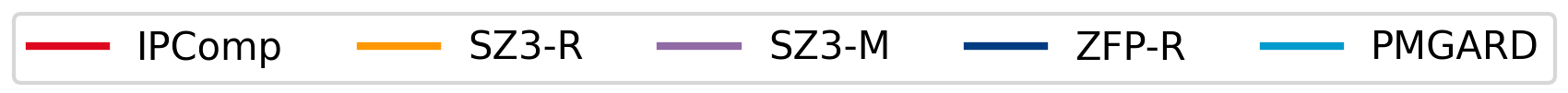}

\subfigure[Density]
{
\raisebox{-1cm}{\includegraphics[scale=0.6]{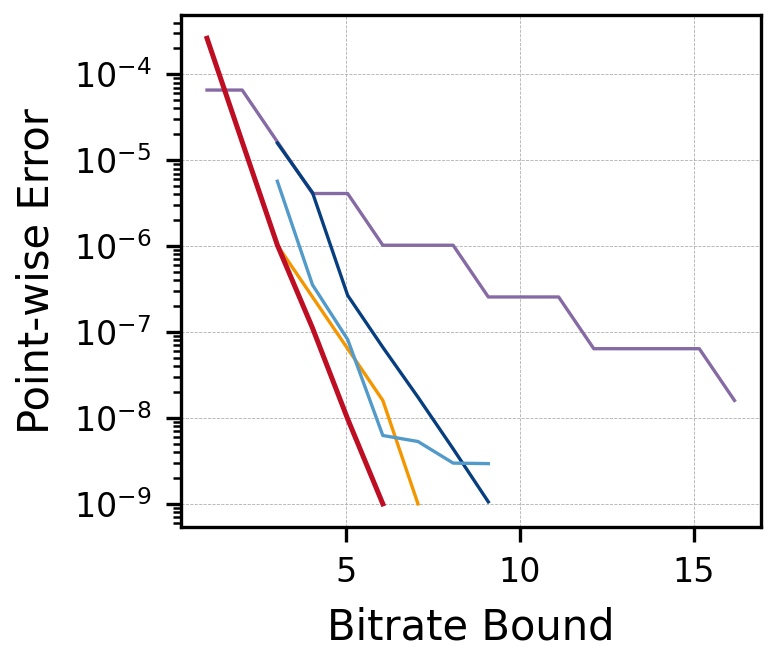}}
}
\subfigure[Pressure]
{
\raisebox{-1cm}{\includegraphics[scale=0.6]{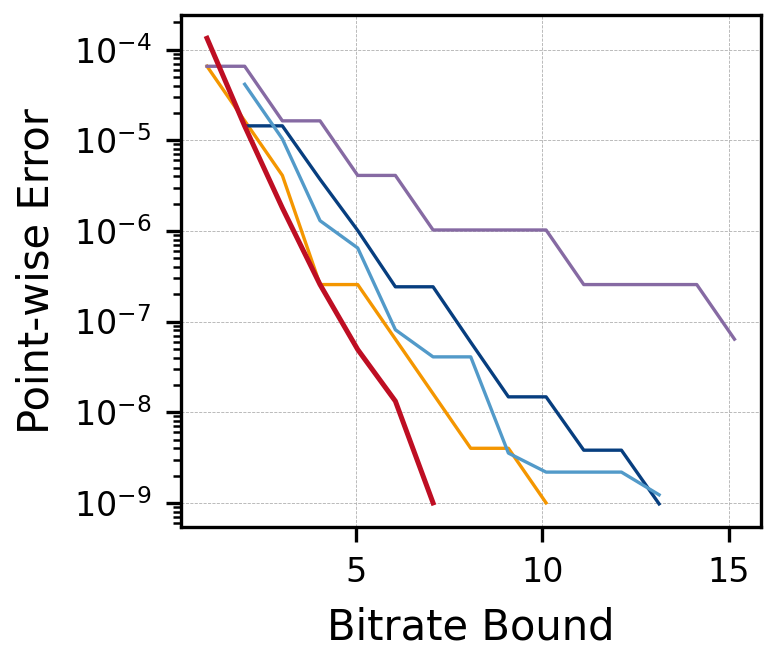}}
}
\vspace{-3mm}

\subfigure[VelocityX]
{
\raisebox{-1cm}{\includegraphics[scale=0.6]{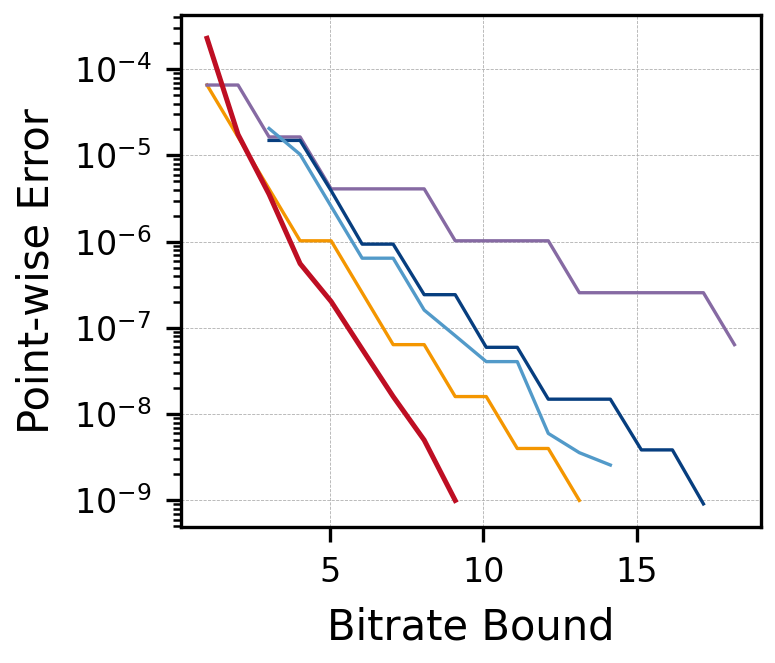}}
}
\subfigure[Wave]
{
\raisebox{-1cm}{\includegraphics[scale=0.6]{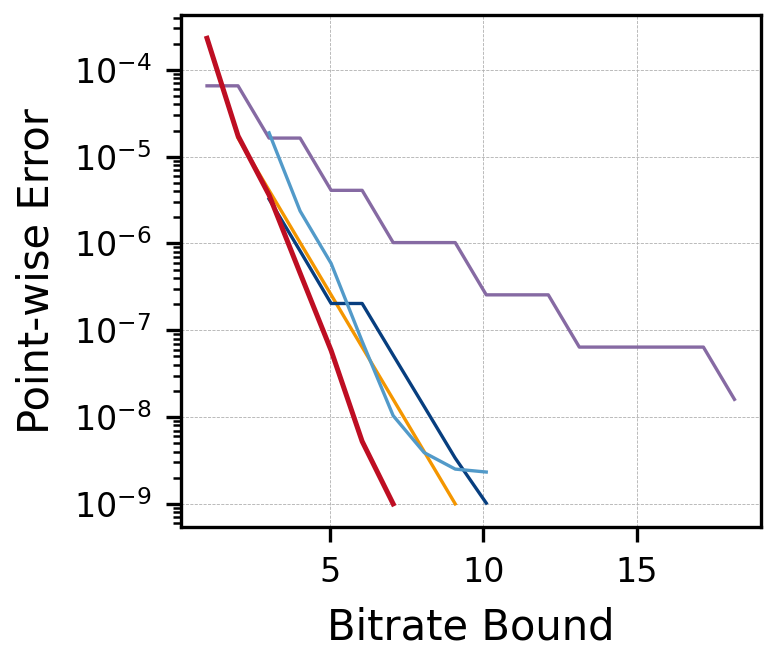}}
}

\vspace{-3mm}

\subfigure[SpeedX]
{
\raisebox{-1cm}{\includegraphics[scale=0.6]{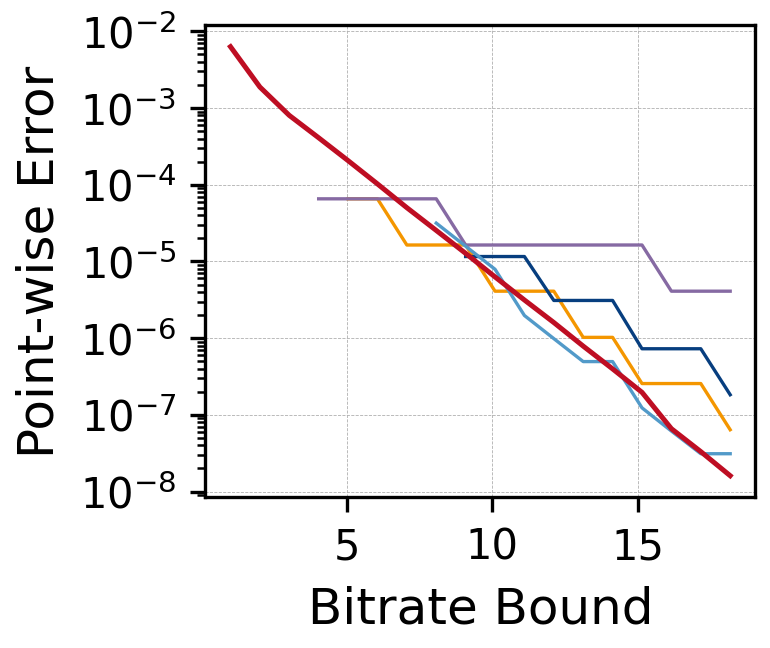}}
}
\subfigure[CH4]
{
\raisebox{-1cm}{\includegraphics[scale=0.6]{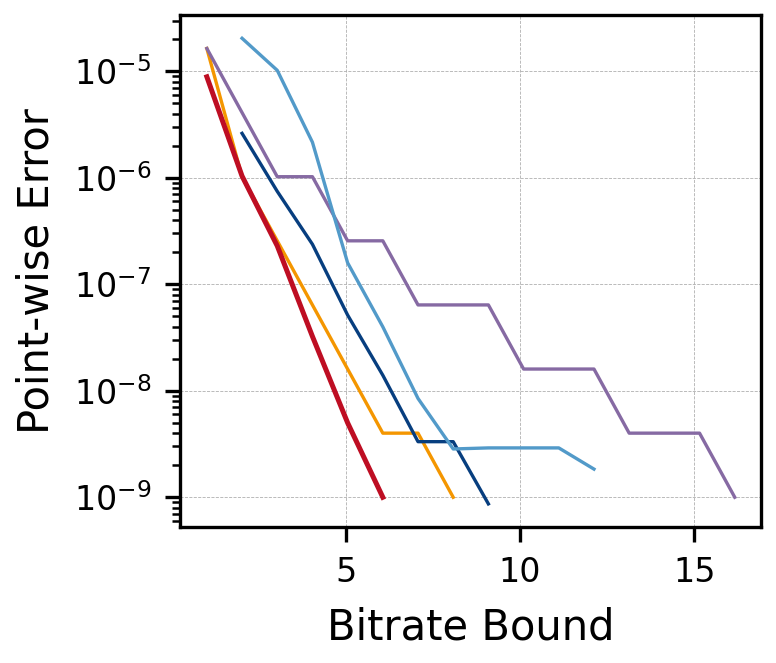}}
}
\caption{Under given bitrate budget, our solution IPComp reconstructs the highest fidelity output than all the baselines (a lower lossy error indicates a high fidelity)}
\label{fig:pwe_bitrate}
\vspace{-3mm}

\end{figure}

\subsubsection{Progressive retrieval efficiency}
After checking the compression ratio, next we evaluate the decompression or retrieval efficiency. Since users may want retrieval by fidelity (specified as error bounds) or size (specified as bitrate), our evaluation is split into two figures for those two scenarios.

\Cref{fig:bitrate_eb} demonstrate the evaluation results of the error bound mode. Compressors with higher efficiency should result in smaller data retrieval size under the error restrictions. The retrieval size is shown as bitrate in~\Cref{fig:bitrate_eb}, so lower lines in the figure indicate better results. 

\Cref{fig:pwe_bitrate} presents the results in fixed bitrate mode. Compressors with higher efficiency should result in higher fidelity (indicated by lower error) with the same retrieval size, still lower lines in the figure indicate better results. 
Since our solution is the only progressive compressor that directly supports the fixed bitrate mode, manual efforts are needed to test other baselines. We enable fix bitrate mode for residual-based compressors by selecting the largest anchor point that fits within the user-specified bitrate constraint. Similarly, for PMGARD, we manually define anchor points ranging from $2^{16}eb$, $2^{15}eb$, $2^{14}eb$ down to $eb$, to allow for bitrate-based decompression.

As shown in both~\Cref{fig:bitrate_eb} and~\Cref{fig:pwe_bitrate}, our compressor IPComp demonstrates its superior compression efficiency by consistently achieving the smallest data load size under the same max error, and the lowest max error under the same bitrate budget.
This advantage primarily comes from the predictive coding method discussed in~\Cref{sec:design coding} and the optimized loading strategy discussed in~\Cref{sec:optimizer}. The residual-based solutions SZ3-R and ZFP-R have a staircase line in both of the figures because the fidelity of such solutions is limited to a few pre-defined residual levels. In comparison, our solution provides higher flexibility as it supports retrieval on arbitrary fidelity and bitrate.




\begin{figure}[ht] \centering
\hspace{1.2mm}
\includegraphics[scale=0.115]{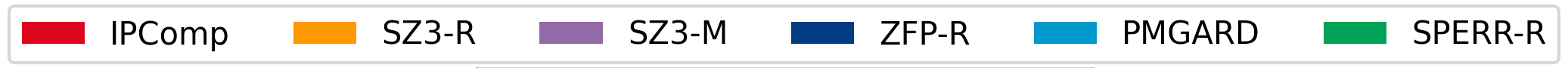}

\hspace{-8mm}
\subfigure[Compression]{
\raisebox{-1cm}{\includegraphics[scale=0.45]{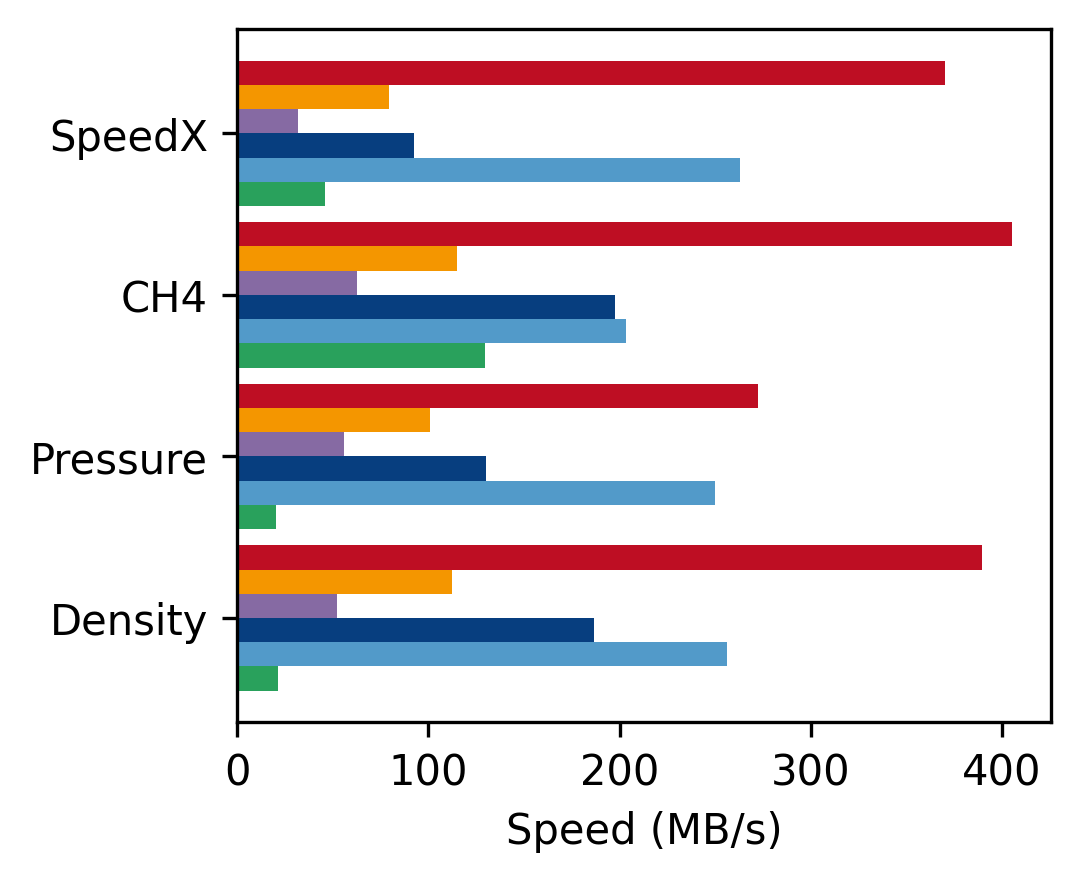}}
}
\subfigure[Decompression]{
\raisebox{-1cm}{\includegraphics[scale=0.45]{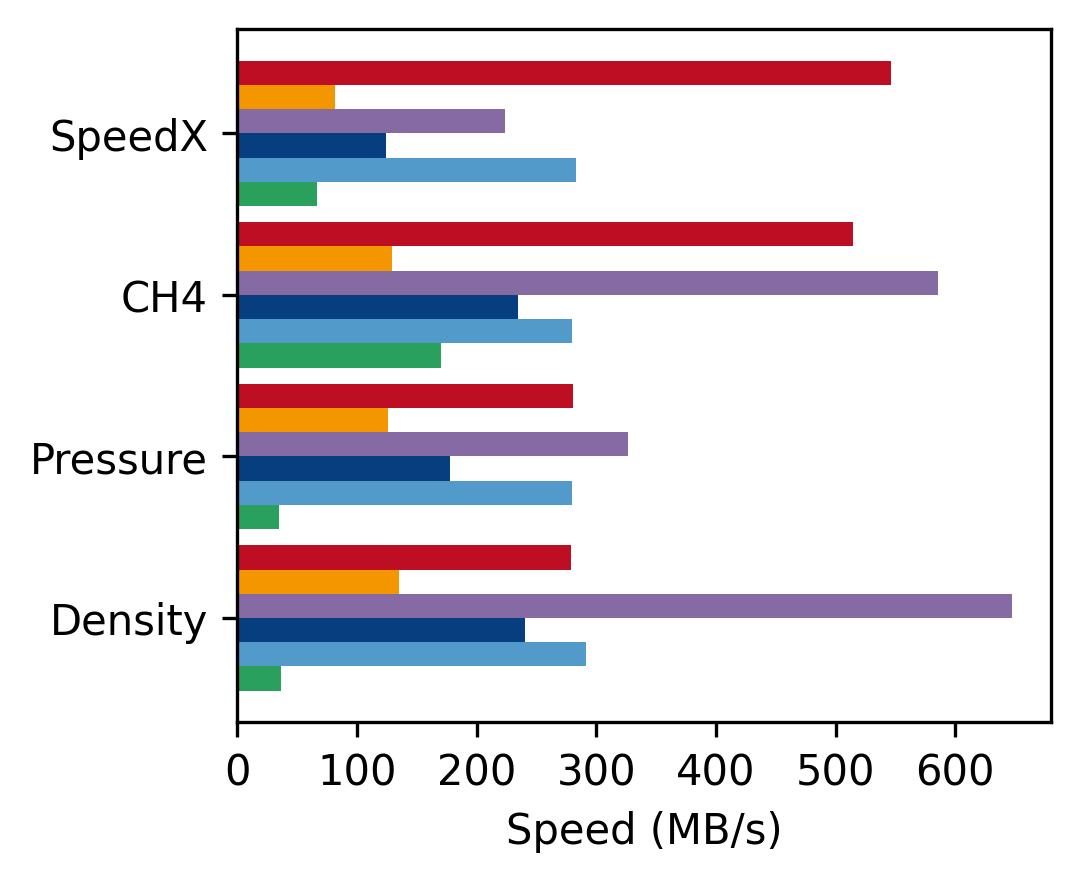}}
}
\caption{Our solution IPComp has the fastest speed in both compression and decompression than all others except for SZ3-M which is multi fidelity but not progressive}
\label{fig:throughput}
\end{figure}

\begin{figure}[ht] \centering
\vspace{3mm}
\hspace{6mm}
\includegraphics[scale=0.12]{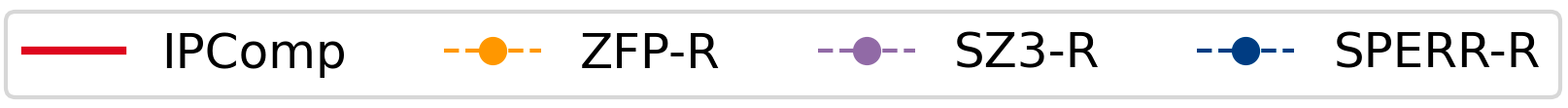}
\subfigure[Compression]{
\includegraphics[scale=0.5]{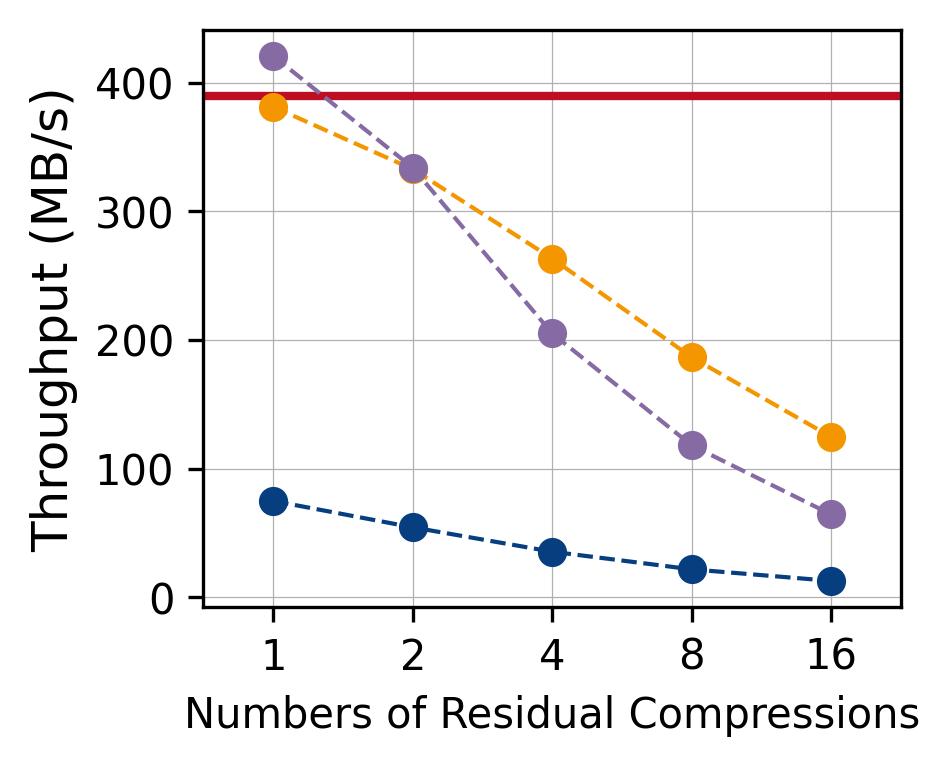}
}
\subfigure[Decompression]{
\includegraphics[scale=0.5]{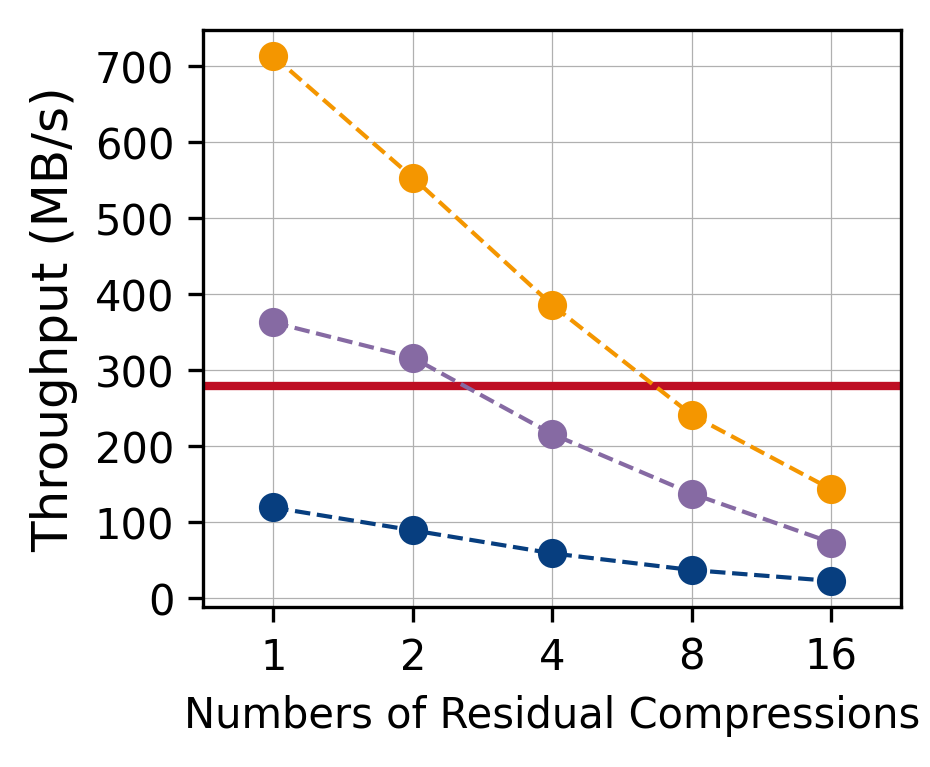}
}
\caption{The speed of residual-based progressive solutions (SZ3-R, ZFP-R, etc) will drop significantly when the setting of residual count increases}
\label{fig:residual}

\end{figure}

\subsubsection{Speed}

We evaluate the compression and retrieval speed of all the compressors. To ensure a fair comparison, we use $eb=10^{-9} \times \text{Range}(dataset)$ to compress and retrieve the data for all compressors, except for PMGARD as it is a lossless compress with lossy retrieval by design. To reach the target $eb$, residual-based compressors need to compress and decompress in multiple iterations.
The speed tests are shown in~\Cref{fig:throughput}. It confirms our solution IPComp is up to around 300\% faster than other approaches in most cases, except for SZ3-M. SZ3-M supports multi fidelity by compressing input in different error bounds and storing all those outputs together. Such a solution is not progressive since the retrieval cannot use previous passes of data. As a result, its compression ratio and retrieval efficiency are both extremely limited in the experiments.
We include one additional compressor SPERR-R in this figure. SPERR-R is the residual progressive implementation of the wavelet-based lossy compressor SPERR~\cite{sperr}. The speed of SPERR-R is extremely slow -- less than 50 MB/s in more than half of the cases. Such a slow speed will offset the savings of reduced data size for I/O, and potentially slow down the scientific workflow. As a result, we do not include SPERR-R as one of our baselines in the full evaluation.

Another evaluation in this section is the speed of residual-based compressors. As discussed in~\Cref{sec: exp-baseline}, residual-based ones must compress and decompress multiple times based on the number of pre-defined error bounds. More pre-defined error bounds would provide more flexibility in retrieval, however, as \Cref{fig:residual} shows, this will instead reduce their speed significantly. We also observe from \Cref{fig:residual} that the residual results are curved instead of straight lines. The reason is that although having more predefined error bounds increases the number of iterations, each iteration takes less time because the looser bounds result in a smaller range of quantized integers. However, the total time still increases significantly due to the cumulative effect of all iterations.
This drawback of residual-based compressors highlights the advantages of our solution which can deliver highly flexible retrieval at high speed.

\begin{figure}[ht] \centering
\includegraphics[scale=0.14]{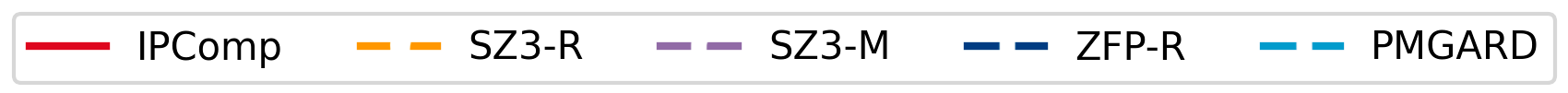}

\hspace{-4mm}
\subfigure[Density]
{
\raisebox{-1cm}{\includegraphics[scale=0.6]{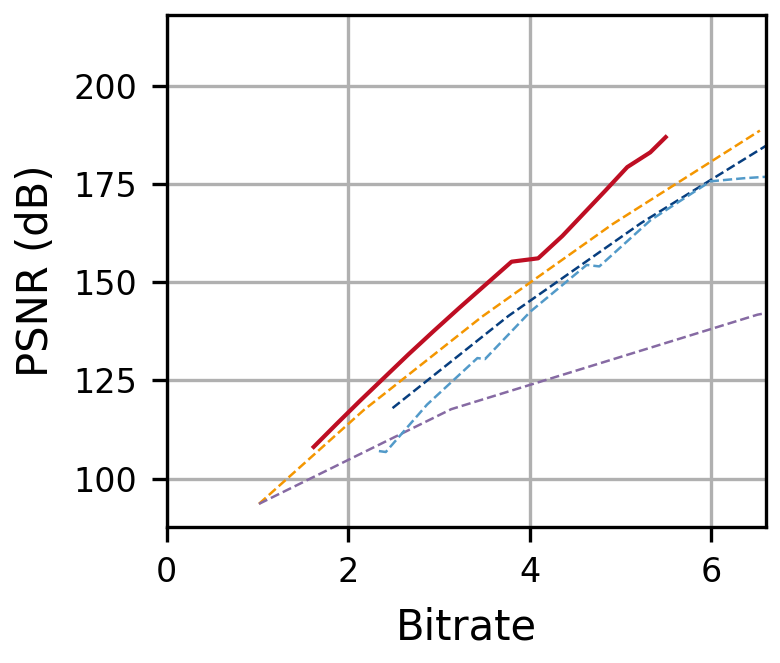}}
}
\subfigure[Pressure]
{
\raisebox{-1cm}{\includegraphics[scale=0.6]{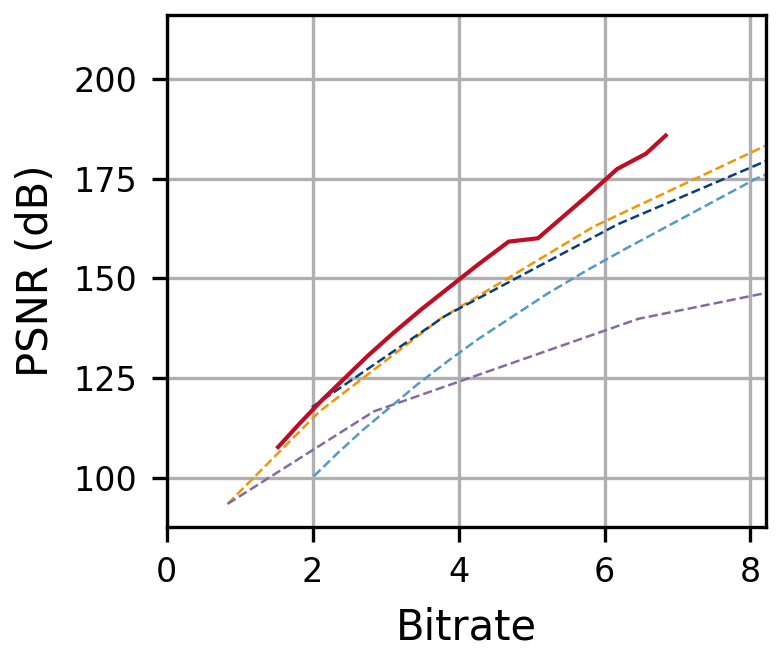}}
}
\vspace{-3mm}

\hspace{-4mm}
\subfigure[VelocityX]
{
\raisebox{-1cm}{\includegraphics[scale=0.6]{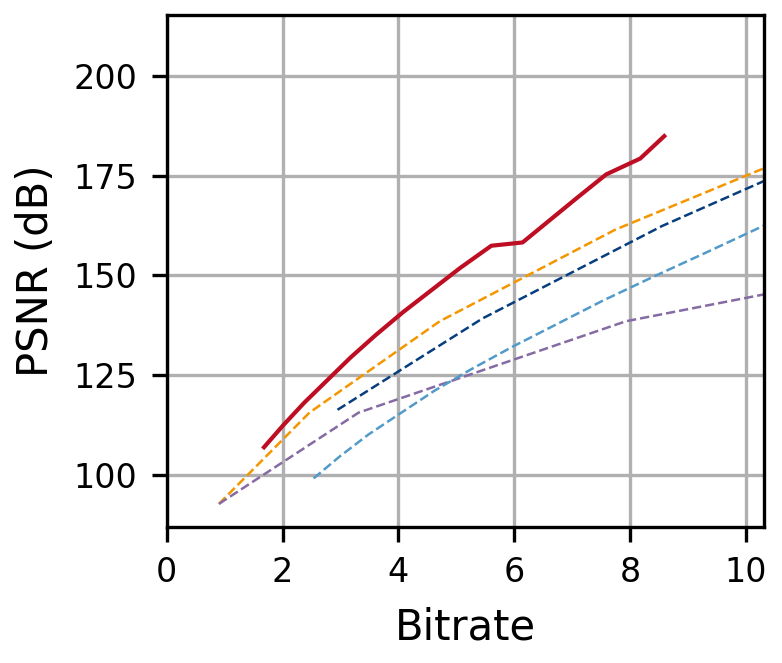}}
}
\subfigure[CH4]
{
\raisebox{-1cm}{\includegraphics[scale=0.6]{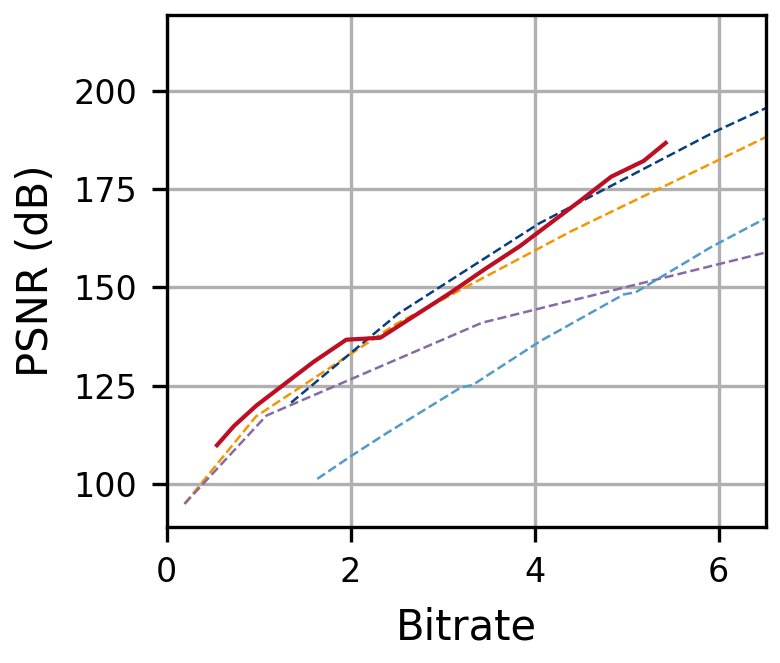}}
}

\vspace{-2mm}
\caption{Our solution leads to higher PSNR under the same bitrate retrieved in most of the cases compared with other solutions}
\label{fig:psnr}
\vspace{-4mm}
\end{figure}

\subsubsection{Progressive retrieval for PSNR}
We also evaluate the PSNR of the compressed data. Since PSNR is mathematically related to the $L_2$ norm error, it serves as another important metric for assessing compression fidelity.
Although our design primarily targets the $L_\infty$ norm and we do not explicitly optimize for PSNR, \Cref{fig:psnr} shows our approach still maintains competitive or superior PSNR compared to other baselines.

\begin{figure}[ht] \centering

\vspace{-4mm}
\subfigure[Curl (0.1\% retrieved)]{
\raisebox{-1cm}{\includegraphics[scale=0.034]{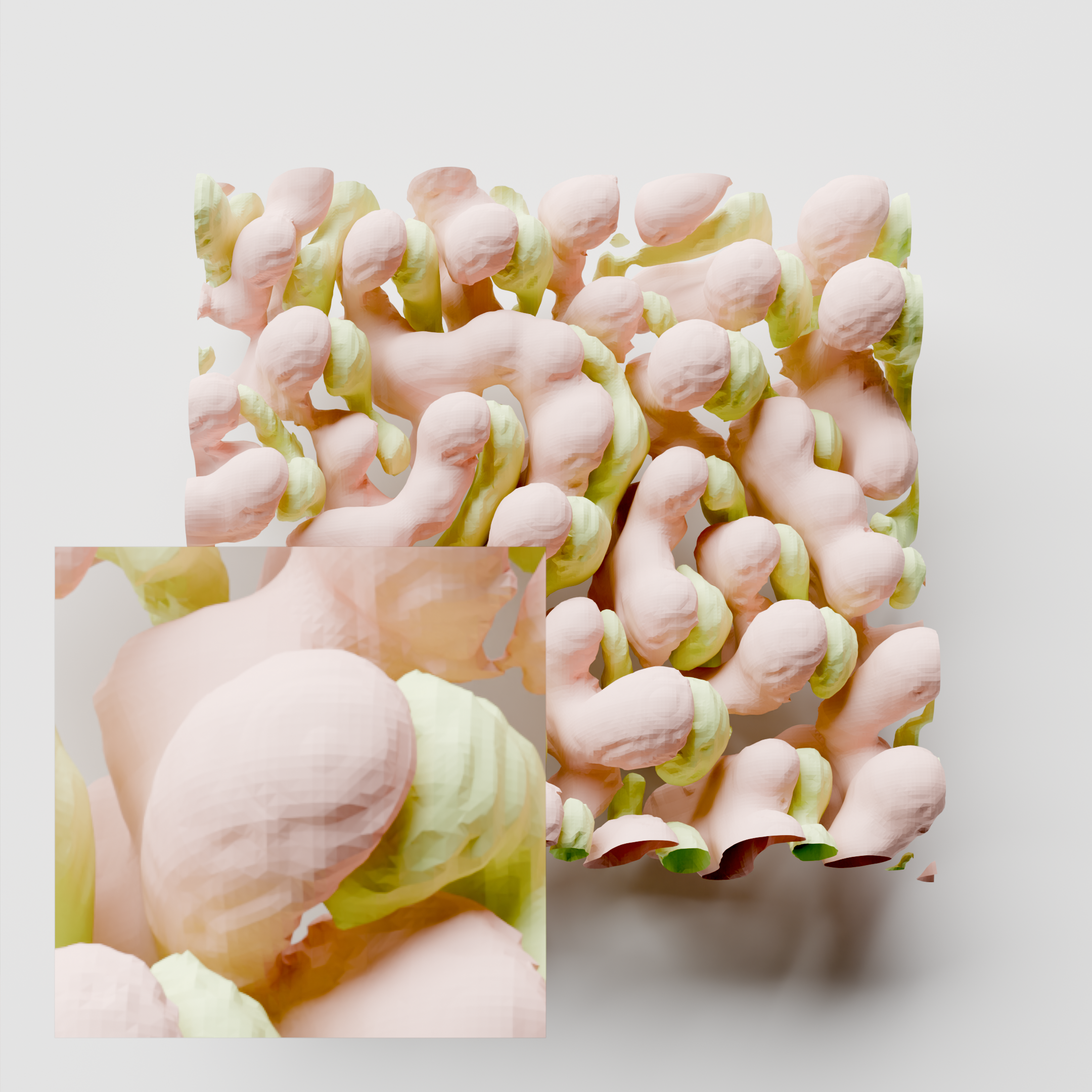}}
}
\subfigure[Curl (0.3\% retrieved)]{
\raisebox{-1cm}{\includegraphics[scale=0.034]{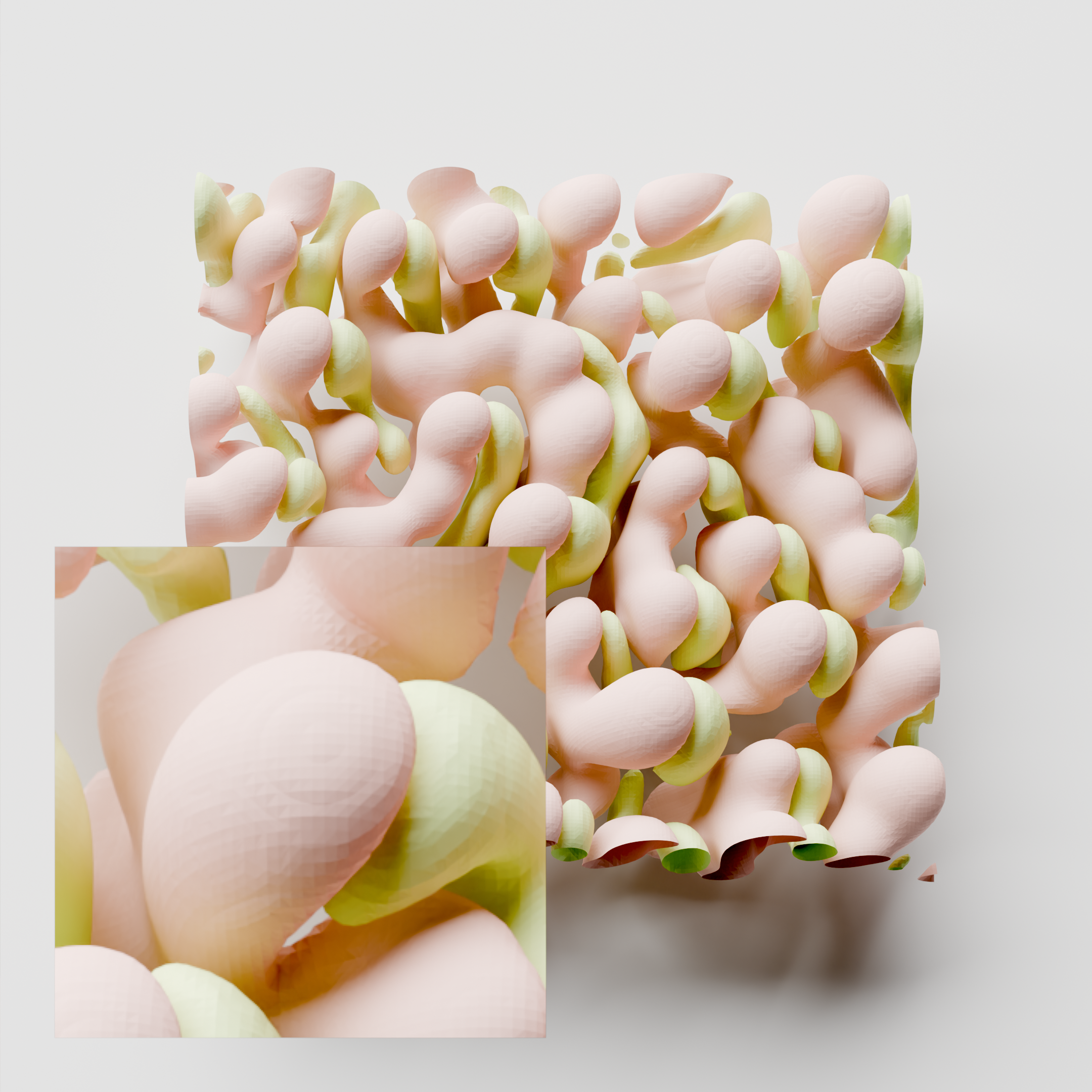}}
}
\subfigure[Curl (1\% retrieved)]{
\raisebox{-1cm}{\includegraphics[scale=0.034]{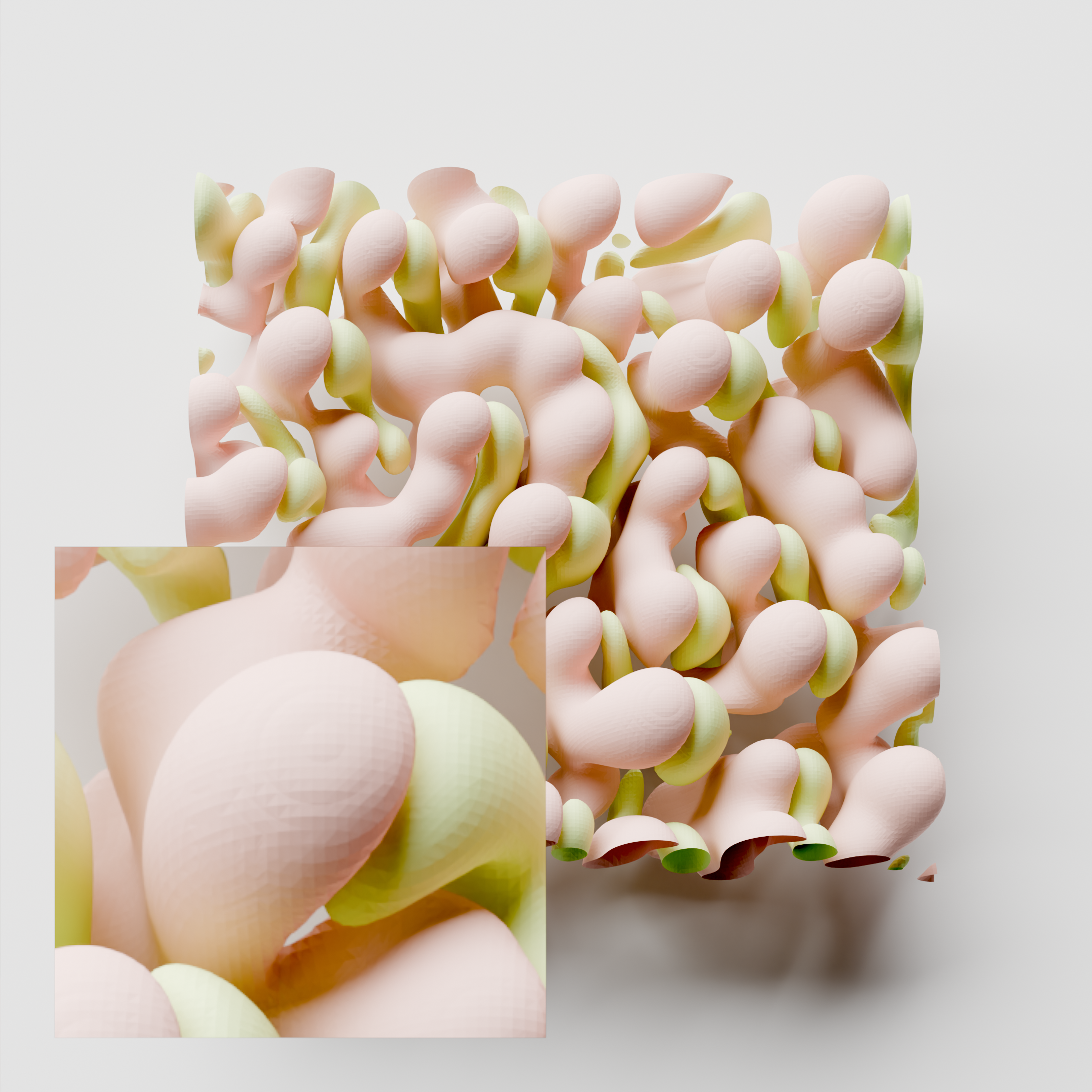}}
}

\subfigure[Laplace (0.1\% retrieved)]{
\raisebox{-1cm}{\includegraphics[scale=0.034]{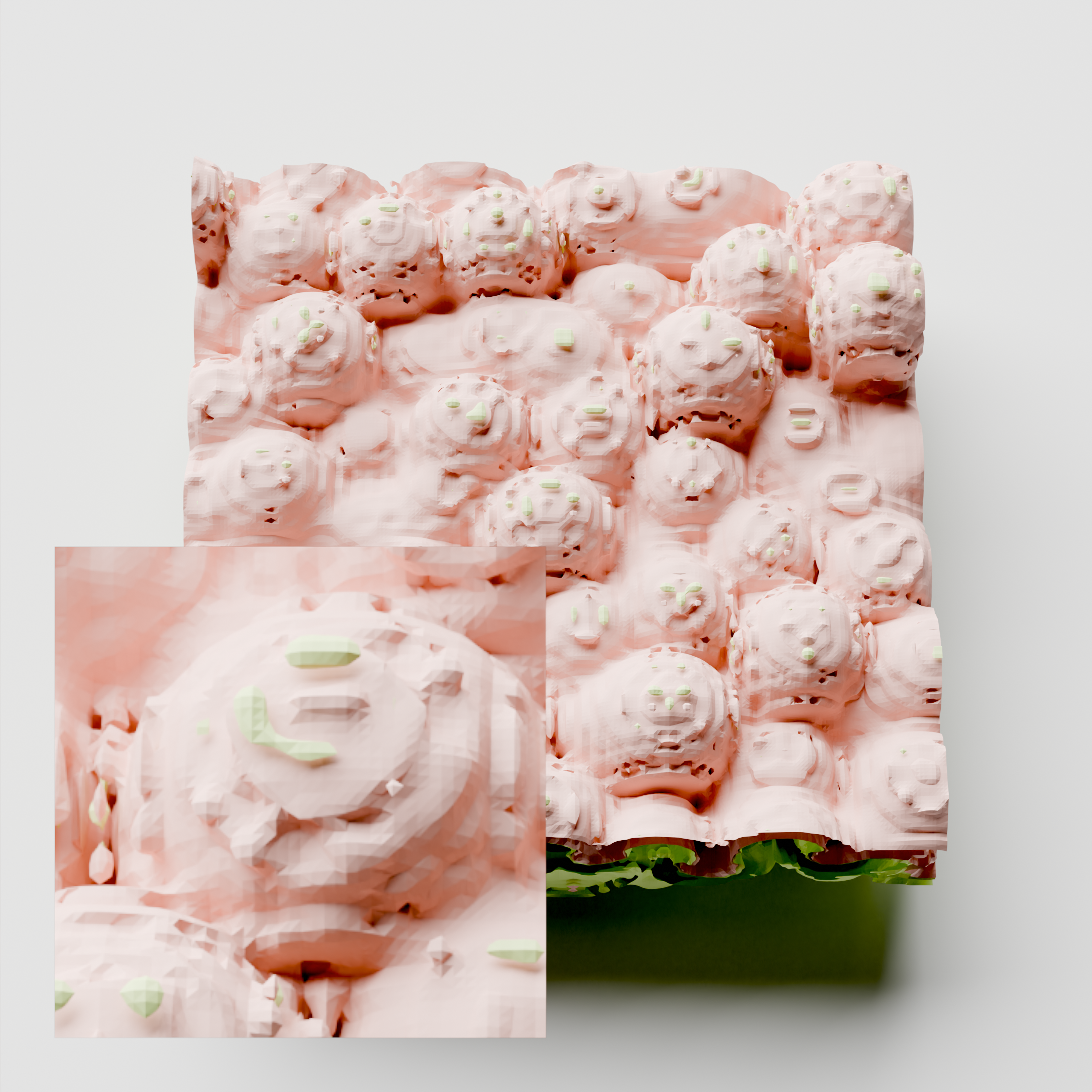}}
}
\subfigure[Laplace (0.3\% retrieved)]{
\raisebox{-1cm}{\includegraphics[scale=0.034]{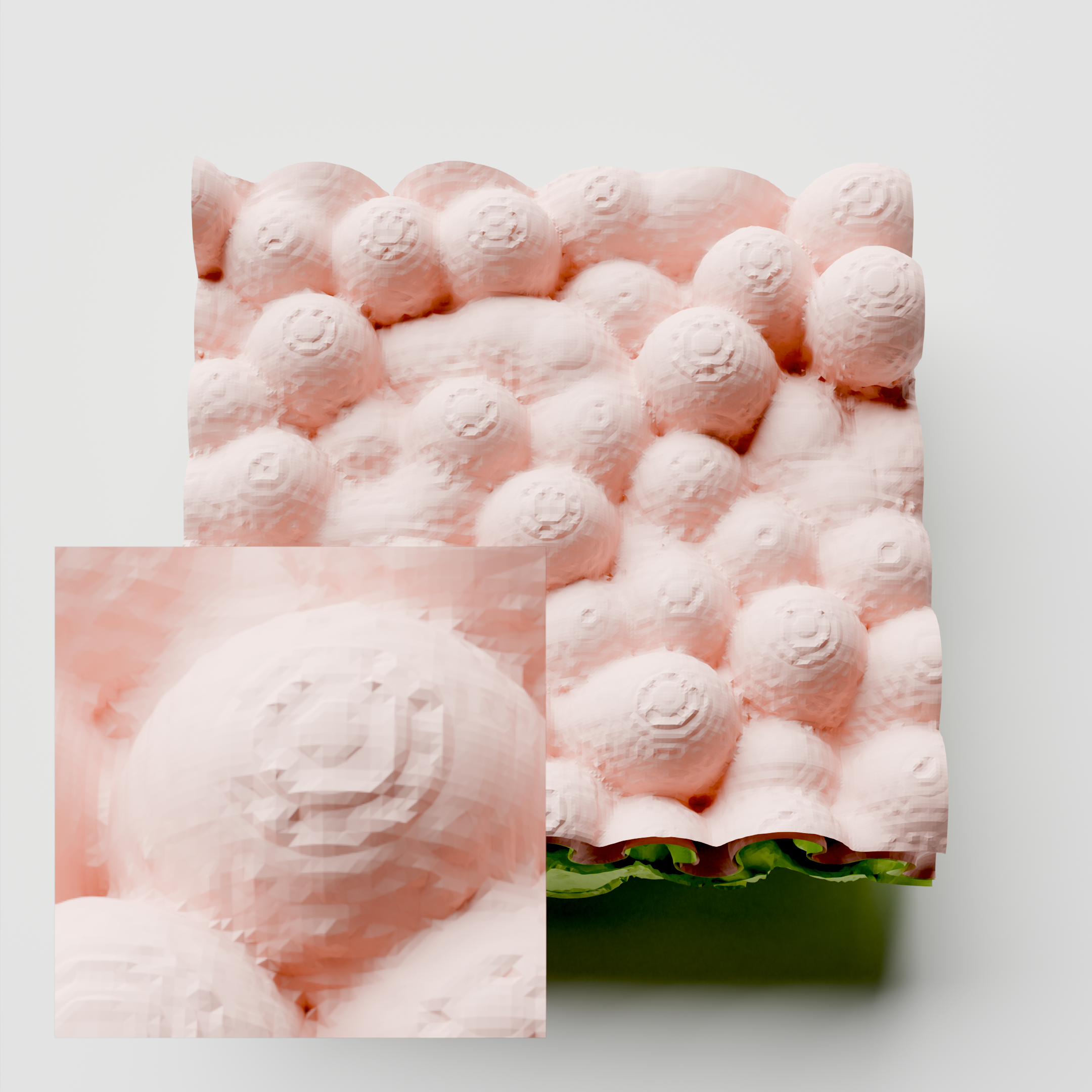}}
}
\subfigure[Laplace (1\% retrieved)]{
\raisebox{-1cm}{\includegraphics[scale=0.034]{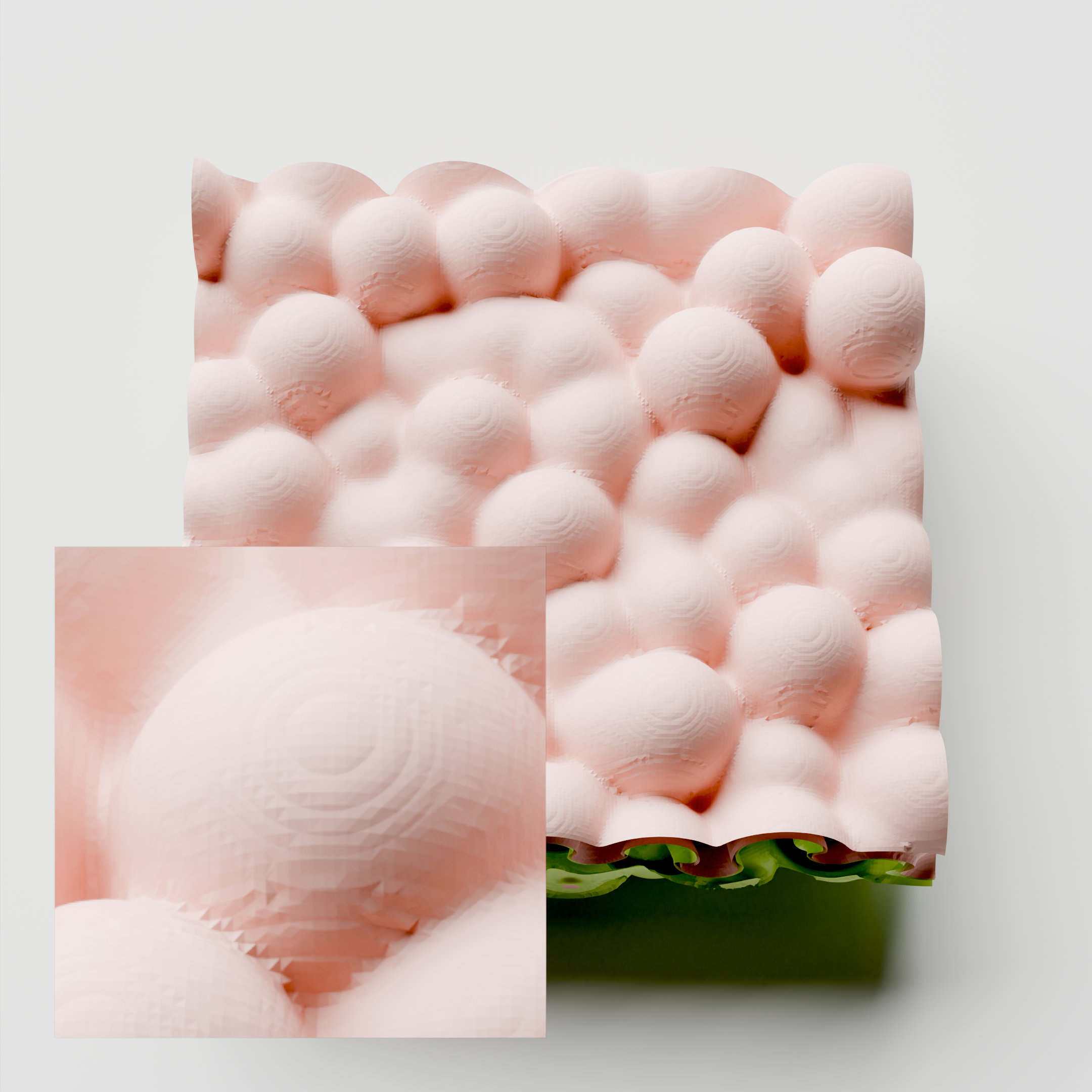}}
}
\vspace{-4mm}
\caption{The visualization of two post-analysis metrics Curl and Laplace on the same Density data. Loading 0.3\% is fine for Curl but Laplace requires 1\% -- demonstrating the necessity of progressive retrieval}
\label{fig:visual}
\vspace{-6mm}
\end{figure}

\subsubsection{Visual quality}
Additionally, we evaluate the visualization quality of the reconstructed data as shown in \Cref{fig:visual}. We load 0.1\%, 0.3\%, and 1.0\% from the same data, and assess the impact on visualization of Curl and Laplacian. While loading 0.3\% of data is enough for Curl in terms of visualization, 1\% of data is needed for Laplacian. This confirms the necessity of progressive retrieval in scientific applications.

\section{Conclusions and Future Work}
\label{sec:conclusion}

In this work, we present IPComp, an interpolation-based progressive lossy compression solution designed to address the growing need for efficient scientific data storage and retrieval. Our approach accomplishes progressiveness effectively with an interpolation prediction model, multi-level bitplanes, and predictive coding techniques. It is equipped with an optimizer to minimize the data volume during retrieval under given error bound or bitrate targets. 

Experimental evaluations conducted on six real-world scientific datasets demonstrate the effectiveness of our solution. IPComp consistently achieves the fastest speed, the highest compression ratios, the lowest data retrieval volume, and the highest data fidelity compared to state-of-the-art alternatives. Additionally, compared with residual-based solutions that only support limited retrieval options, our approach is very flexible on fidelity control as it takes arbitrary error bounds and bitrates as retrieval options. 

Our findings suggest that IPComp represents a significant advancement in progressive lossy compression and is a practical choice for scientific applications. The future work will focus on optimizing hardware acceleration (e.g., GPU and tensor cores), integrating with scientific workflows like HDF5, and expanding large-scale HPC evaluations. These improvements will further refine IPComp for broader application scenarios.

\section{Acknowledgments}
This research was supported by the National Science Foundation under Grant OAC-2104023, OAC-2311875, OAC-2311876, OAC-2311878, OAC-2344717, and by the U.S. Department of Energy, Office of Science, Advanced Scientific Computing Research (ASCR), under contract DE-AC02-06CH11357.
This work used the Purdue Anvil CPU cluster through allocation CIS230308 and CIS240192 from the Advanced Cyberinfrastructure Coordination Ecosystem: Services \& Support (ACCESS) program.

\bibliographystyle{ACM-Reference-Format}
\bibliography{citations.bib}

\end{document}